\documentclass[graybox]{svmult}
\usepackage{type1cm}        
\usepackage{makeidx}         
\usepackage{graphicx}        
\usepackage{multicol}        
\usepackage[bottom]{footmisc}

\usepackage{newtxtext}       %
\usepackage[varvw]{newtxmath}       

\makeindex             

\usepackage{latexsym}
\usepackage{amsfonts,euscript}
\usepackage{sidecap}
\usepackage{color}
\usepackage{amsmath, empheq} 
\usepackage{newtxmath}  
\usepackage{wrapfig}
\usepackage{graphicx}
\usepackage{caption}
\usepackage{subcaption}

\newcommand{\bs}{\begin{eqnarray*}}
\newcommand{\es}{\end{eqnarray*}}

\newcommand{\bu}{\bf u}

\newcommand{\bmat}{\begin{pmatrix} }
\newcommand{\emat}{\end{pmatrix}}
\newcommand{\beqs}{\begin{eqnarray*}}
\newcommand{\eeqs}{\end{eqnarray*}}
\newcommand{\beq}{\begin{eqnarray}}
\newcommand{\eeq}{\end{eqnarray}}

\newcommand{\pr}{\partial}
\newcommand{\field}[1]{\mathbb{#1}} 
\newcommand{\B}{\mathbf}
\newcommand{\bep}{\pmb{\epsilon}}

\newcommand{\im}{{\rm Im }\, }
\newcommand{\re}{{\rm Re }\, }

\newcommand{\mf}{\mathfrak}
\newcommand{\bx}{\B{x}}

\usepackage{float}
\usepackage{wrapfig}
\usepackage{circuitikz}
\usepackage{xcolor}
\usepackage{graphicx}
\usepackage[normalem]{ulem}
\usepackage{soul}
\usepackage{hyperref}
\hypersetup{
    colorlinks=true,
    linkcolor=blue,
    filecolor=magenta,      
    urlcolor=cyan,
}
\newcommand{\ov}{\overline}
\newcommand{\wt}{\widetilde}
\colorlet{colorYO}{red}
\colorlet{colorAL}{blue!50}

\newcommand{\bmu}{\pmb{\mu}}
\newcommand{\bo}{{\mathbf 0}}
\usepackage{stmaryrd}
\newcommand{\bn}{{\mathbf n}}
\newcommand{\bpi}{\pmb{\pi}}
\newcommand*{\eqdef}{=\mathrel{\vcenter{\baselineskip0.5ex \lineskiplimit0pt
			\hbox{\scriptsize.}\hbox{\scriptsize.}}}%
}
\def\C{\mathbb{C}}
\def\R{\mathbb{R}}
\newcommand{\ep}{\epsilon}
\newcommand{\bK}{{\mathbf K}}
\newcommand{\by}{{\mathbf y}}
\newcommand{\bI}{{\mathbf I}}
\newcommand{\be}{{\mathbf e}}
\newcommand*{\defeq}{\mathrel{\vcenter{\baselineskip0.5ex \lineskiplimit0pt
			\hbox{\scriptsize.}\hbox{\scriptsize.}}}%
	=}
\newcommand{\bv}{{\mathbf v}}	
\newcommand{\bA}{{\mathbf A}	}
\newcommand{\bb}{{\mathbf b}}
\newcommand{\norm}[1]{\left\lVert#1\right\rVert}
\newcommand{\mc}{\mathcal}
\usepackage{dutchcal}
\DeclareMathAlphabet{\pazocal}{OMS}{zplm}{m}{n}

\begin{document}
\title*{On applications of Herglotz-Nevanlinna functions in material sciences{, II: extended applications and  generalized theory }}
\titlerunning{Herglotz-Nevanlinna functions: II}
\author{Miao-Jung Yvonne Ou and Annemarie Luger}
\institute{Miao-Jung Yvonne Ou\at Department of Mathematical Sciences, University of Delaware, Newark, DE 19716, USA, \email{mou@udel.edu}
\and Annemarie Luger\at  Department of Mathematics, Stockholm University, SE-106 91 Stockholm, Sweden\email{ luger@math.su.se}}
%
%
\maketitle

\abstract{Part II of the review article focuses on the applications of Herglotz-Nevanlinna functions in material sciences. It {presents} a diverse set of applications with {details and} the role of Herglotz-Nevanlinna functions clearly pointed out. {This paper is concluded by} a collection of existent {generalizations of the class of Herglotz-Nevanlinna functions} that are motivated by potential applications.}

\section{\label{intro}Introduction}
{In this part of the review paper, we present a wide class of applications of Herglotz-Nevanlinna functions in material sciences. We start with the application in the static theory of two-phase composite materials, where the scalar-valued Herglotz-Nevanlinna functions correspond to the effective properties of the composite materials. Following this is an example showing how the matrix-valued Herglotz-Nevanlinna function theory can be applied to study the permeability tensor of a porous material. In both applications, the independent variable of the corresponding Herglotz-Nevanlinna functions is the contrast of material properties. The other group of applications presented in this paper demonstrates the power of Herglotz-Nevanlinna functions in the study of systems of equations where the energy dissipation and dispersion satisfied causality and passivity, whose mathematical definition can be clearly specified in terms of the Herglotz Nevanlinna functions. After presenting their various applications in material sciences, we conclude this paper by introducing several classes of functions which can be considered as various generalizations of the Herglotz-Nevanlinna functions motivated by some emerging research field in physics and engineering.} 

{The paper is organized as follows.} Section \ref{subsec:composite} deals with composite materials  and bounds on effective properties. 
In Section \ref{subsec:memory} it is demonstrated how the usage of Herglotz-Nevanlinna functions can avoid a numerically costly memory term in the modeling of materials.  Section \ref{subsec:milton} shows how the bounds for quasi-static cloaking can be derived. In Section \ref{subsec:Figotin} a general representation theorem of Herglotz-Nevanlinna functions is used in order to identify certain time dispersive and dissipative systems as restrictions of Hamiltonian systems. 

Even if all these examples demonstrate the effectiveness of Herglotz-Nevanlinna functions, there are situations in applications that cannot be treated by these methods, but would require more general classes of functions. This applies for instance  for  non-passive systems appearing in electromagnetics, for which  the analytic function in question might have non-positive imaginary part as well.  Another example are  composite materials with more than two phases. Then, even if the corresponding analytic functions still have positive imaginary part, they are not covered by the treatment above, since they depend on more than one complex variable. 

In {Section} \ref{sec:general} we therefore provide an overview of the mathematics that is available for different classes of functions that extend the classical Herglotz-Nevanlinna class and are expected to be relevant for applications in material science. 

{Note that items that are already defined in Part I will not be defined again in this part. }

We hope that this review  can be both helpful for people working in applications (by providing mathematical references for different aspects of Herglotz-Nevanlinna functions as well as their generalizations for future work) and  interesting for pure mathematicians (by pointing out some relevant applications of Herglotz-Nevanlinna functions).

\section{Applications}
{This section starts with the applications arising in the study of effective properties of composite materials, followed by the application in broadband passive quasi-static cloaking and is concluded with a delicate application of the operator-valued Herglotz-Nevanlinna function theory for understanding  the Hamiltonian structure of  time-dispersive and dissipative systems.  }

\subsection{Effective properties of two-phase composite materials}\label{subsec:composite}

\subsubsection{Effective properties of composite materials and bounds by using theory of the Stieltjes function}

Composite materials made of pure homogenous phases are abundant around us, eg. reinforced concrete, plywood, fluid saturated sand, cancellous bones and sea-ice. Suppose the scale of the microstructure of a bulk composite sample is much smaller than the size of the sample{;} it makes sense to use the effective moduli  to describe the properties of composite materials. For example, the effective permittivity of a complex fluid or the effective Young's modulus of a cancellous bone sample. Intuitively, these effective properties should depend on the properties of the pure phases as well as how these constituents are arranged, i.e. the microstructure of the composite. For multi-laminated microstructure, there are exact algebraic {formulas} for computing the effective properties as certain averages of the properties of the constituents{;} see \cite{Milton2002The-Theory-of-C}. However, for most microstructures, there is no exact "mixing theory formula" that can be used to compute the effective properties even though the effective properties are well defined by the homogenization theory \cite{Papanicolaou2011Asymptotic-Anal}\cite{Jikov1994Homogenization-}\cite{zhikov2014introduction}\cite{Tartar2010The-General-The}. For the history of the development and the limitations of various formulas for computing the effective dielectric constants for simple microstructures, see \cite{BERGMAN1978377} for details. Instead of looking for the exact formulas, many researchers have looked into the possibility of finding bounds of effective properties from the given constituent properties and information of the microstructure{; see } \cite{Wiener1912Die-theorie-des}\cite{HASHIN1962343,HASHIN1963127,Hashin1962A-Variational-A}\cite{prager1969improved-variat} \cite{BERGMAN1978377}\cite{PhysRevLett.44.1285}\cite{bergman1982rigorous-bounds}\cite{PhysRevB.23.3058}\cite{milton1981bounds-on-the-c}\cite{Milton1981Bounds-on-the-t}\cite{McPhedran1981Bounds-and-exac}\cite{golden1983bounds-for-effe}\cite{dellantonio1986an-approach-thr}\cite{milton1990representations}\cite{Bruno1990Interchangeabil}\cite{GOLDEN1986333}, just to name a few. From this vast and rich literature emerges the beautiful bounding method based on the analytic properties {of} the effective moduli as a Stieltjes function of the dielectric constants of the pure phases; it was first  described in \cite{BERGMAN1978377} by David Bergman and further developed and extended from real-valued bounds to the general complex bounds by Graeme Milton in \cite{Milton1980Bounds-on-the-c}\cite{milton1981bounds-on-the-c}. This method provides a way for deriving the bounds without the use of variational principles. The first rigorous derivation of the Stieltjes function representation for the effective dielectric parameters of a two-phase composite is given in 1983 by Kenneth Golden and George Papanicolaou \cite{golden1983bounds-for-effe} in a random media setting. To fix ideas, we start with a brief description of the proof in \cite{golden1983bounds-for-effe}. 

Let $(\Omega,\mathcal{F},P)$ be a probability space and the permittivity tensor $\bep(\bx,\omega)$ a stationary random field, where $\omega$ is a realization in $\Omega$ and $\bx$ the spatial coordinates in $\field{R}^d$ with $d\in \field{N}$ and $d\ge 2$. Specifically, there exists a bijective group transformation $\pmb{\tau}_\bx$ from $\Omega$ to $\Omega$, $\pmb{\tau}_\bx\pmb{\tau}_\B{y}=\pmb{\tau}_\B{x+y}$ for all $\bx,\B{y}\in \field{R}^d$ such that $P(\pmb{\tau}_\bx A)=P(A)$ for all $\bx\in \field{R}^d$ and $A\in \mathcal{F}$. Suppose the permittivity tensor $\bep(\bx,\omega)$ can be represented by a  measurable function $\tilde{\bep}(\omega)$ on $\Omega$ as follows
\beq
\bep(\bx,\omega)=\tilde{\bep}(\pmb{\tau}_{-\bx} \omega){.}
\label{stationary}
\eeq
It is further assumed to be bounded and satisfies the ellipticity condition, i.e. there exist two positive numbers $\alpha$ and $\beta$ so that $\alpha \pmb{\xi}\cdot \pmb{\xi} \le \bep(\bx,\omega)\pmb{\xi}\cdot \pmb{\xi} \le \beta \pmb{\xi}\cdot \pmb{\xi}$ for all $\bx,\pmb{\xi}\in \field{R}^d$. Since all the random fields considered here are stationary, the solutions of the form in \eqref{stationary} are sought for, i.e.
\beq
\B{E}(\bx,\omega)=\tilde{\B{E}}(\pmb{\tau}_{-\bx} \omega),\,\B{D}(\bx,\omega)=\tilde{\B{D}}(\pmb{\tau}_{-\bx} \omega){.}
\label{sta_solutions}
\eeq
Consider the electrostatic Maxwell's equations for the random stationery electric field $\B{E}(\bx,\omega)$ and the electric induction field $\B{D}(\bx,\omega)$ 
\beq
\B{D}(\bx,\omega)=\bep(\bx,\omega) \B{E}(\bx,\omega),\, \nabla\times \B{E}(\bx,\omega)=\B{0},\, \nabla\cdot \B{D}(\bx,\omega)=0\nonumber\\ \mbox{ and } \int_\Omega  \B{E}(\bx,\omega) P(d\omega)=\overline{\B{E}}
\label{static}
\eeq
with a prescribed constant electric field $\ov{\B{E}}$ as its assemble average. {Let} the constant vector $\ov{E}$ be $\B{e}_j$, the unit vector in the $j$-th direction, $j=1,\cdots,d$ and denote the corresponding solution as $\B{E}^j$ and $\B{D}^j$. The effective permittivity tensor $\bep^*$ is then  defined as 
\beq
\bep^* \B{e}_l:= \int_\Omega  \B{D}^l(\bx,\omega) P(d\omega)=\int_\Omega \bep(\bx,\omega)\B{E}^l(\bx,\omega) P(d\omega) ,\, l=1,\cdots d.
\label{G_eff_def}
\eeq
It can be shown that these assemble averages of the solution do not depend on $\bx$. 
The variational formulation plays an important role in the proof; it is described here. First consider the Hilbert space $H:=L^2(\Omega,\mathcal{F},P)$
 endowed with the inner product $(\tilde{f},\tilde{g})_H:=\int_\Omega \tilde{f}(\omega)\tilde{g}(\omega) P(d\omega)$. Define the operator $T_\bx$ acting on $\tilde{f}\in H$ as $T_\bx \tilde{f}(\omega):=\tilde{f}(\pmb{\tau}_{-\bx} \omega)$. Because $\pmb{\tau}_\bx$ is measure preserving, $T_\bx$ forms a unitary group and has closed densely defined infinitesimal generator $L_j:=\frac{\partial}{\partial x_j} T_\bx\Big|_{\bx=0}$ for each $j=1,\cdots,d$ with domain $\mathcal{D}_j\subset H$. 
 Then $\mathcal{D}:=\bigcap_{j=1}^d \mathcal{D}_j \subset H$ is a Hilbert space with the inner product $(\tilde{f},\tilde{g})_{\mathcal{D}}:=\int_\Omega \tilde{f}(\omega)\tilde{g}(\omega) P(d\omega)+\sum_{i=1}^d\int_\Omega L_i \tilde{f}(\omega) L_i\tilde{g}(\omega) P(d\omega)$. 

Since the problem \eqref{static} is equivalent to finding $\B{E}$ in a curl-free space, and $\B{E}=\B{\ov{E}}+\B{G}$ with a zero-average field $\B{G}$, the following Hilbert space {of vector-valued functions}  with inner product $(\cdot,\cdot)_\mathcal{H}:=(\cdot,\cdot)_{H}$ is considered
\[
\mathcal{H}:=\left\{\tilde{f}_j(\omega)\in H | L_i\tilde{f}_j=L_j\tilde{f}_i \mbox{ weakly },\, \int_\Omega \tilde{f}_j(\omega)P(d\omega)=0,\, i, j=1,\cdots,d \right\}{.}
\]
The variational formulation of \eqref{static}, after taking into account \eqref{sta_solutions}, is to find $\widetilde{\B{G}}^l(\omega)\in \mathcal{H}$ such that 
\beq
\int_\Omega \wt{\bep}(\omega)(\wt{\B{G}}^l(\omega)+\B{e}_l)\wt{\B{f}}(\omega) P(d\omega)=\B{0}\mbox{ for all } \wt{\B{f}}\in \mc{H}.
\label{Golden_var}
\eeq
{Recall that $\B{e}_l$ is the unit vector in the $l$-th direction, $l=1,\cdots,d$.}
This problem is well-posed because the bilinear form is coercive w.r.t. the $H$-norm and the Lax-Milgram lemma can be applied. Letting $\wt{\B{f}}=\wt{\B{G}}^k$ in \eqref{Golden_var} and using the definition in \eqref{G_eff_def} and the fact that $\B{e}_k=\wt{\B{E}}^k-\wt{\B{G}}^k$ by definition, one obtains the following symmetric form  
\beq
\bep^*\B{e}_l \cdot \B{e}_k =\epsilon^*_{kl}=\int_\Omega \wt{\bep}(\omega)\wt{\B{E}}^l(\omega) \wt{\B{E}}^k(\omega)P(d\omega), k,l=1,\cdots,d
{.}
\eeq
To specialize to {the} two-phase case with isotropic constituents, consider $\bep(\bx,\omega)=\chi_1(\bx,\omega)\epsilon_1\B{I}+\chi_2(\bx,\omega)\epsilon_2\B{I}$ with $0<\epsilon_1\le\epsilon_2<\infty$ and indicator functions $\chi_p$, $p$=1,2 such that $\wt{\chi_1}(\omega)+\wt{\chi_2}(\omega)=1$. For example, $\wt{\chi_1}(\omega)$ equals one for all realizations $\omega\in\Omega$ for which the origin is occupied by the material with permittivity $\epsilon_1$.  Define the contrast $h:=\frac{\epsilon_2}{\epsilon_1}$. The variational formulation now reads
\beq
\int_\Omega  [\wt{\chi_1}(\omega)+h \wt{\chi_2}(\omega)](\wt{\B{G}}^l(\omega)+\B{e}_l)\cdot \wt{\B{f}}(\omega) P(d\omega)=\B{0}\mbox{ for all } \wt{\B{f}}\in \mc{H}
\label{Golden_var_h}
\eeq
and the effective permittivity $\bep$ defined in \eqref{G_eff_def}, as a function of $h$, can be expressed as follows 
\beq
\epsilon^*_{kl}(h)=
\epsilon_1\left[\delta_{kl}+(h-1)\int_\Omega(\wt{\chi_2}(\omega)\wt{E}^l_k(h,\omega)P(d\omega)\right]
\label{eff_perm_h}
\eeq
or equivalently, one can focus on the function 
\beq
m_{kl}(h):=(\epsilon_1)^{-1}\epsilon^*_{kl}(h)=\left[\delta_{kl}+(h-1)\int_\Omega(\wt{\chi_2}(\omega)\wt{E}^l_k(h,\omega)P(d\omega)\right]
\label{m_function}
\eeq
{Note that $m_{kl}(1)=\delta_{kl}$ by definition.}
If one replaces the inner product $(\cdot,\cdot)_{H}$ with the one for complex-valued functions (by complex conjugating one of the functions), then the sesquilinear form in \eqref{Golden_var_h} is {coercive} in $h\in\field{C}\setminus (-\infty,0]$. Hence by Lax-Milgram lemma,  there is a unique solution for all $\epsilon_1,\epsilon_2\in \field{C}$ such that $\frac{\epsilon_2}{\epsilon_1}\in\field{C}\setminus (-\infty,0]$. This further implies that $\bep(h)$ is analytic in $h\in\field{C}\setminus (-\infty,0]$ and so is the effective permittivity $\bep^*(h)$. 

To obtain the spectral representation, first note that \eqref{Golden_var_h} can be written formally in terms of {the} Kronecker $\delta$ as (by thinking of $\wt{G}_k^l$ as the gradient of a function because it is curl-free, i.e. $\wt{G}_k^l=L_k\psi^l$ {for some scalar function $\psi^l$}, $k=1,\dots,d$)
\beqs
\sum_{k=1}^d L_k[\wt{\chi_1}(\omega)+h\wt{\chi_2}(\omega)(\wt{G}_k^l+\delta_{kl})]=0, l=1,\cdots,d.
\eeqs
Rewriting the above expression {in} the following form 
\beq
\sum_{k=1}^d L_k  \wt{G}_k^l+ (h-1) \sum_{k=1}^d  L_k \wt{\chi_2}(\omega)(\wt{G}_k^l+\delta_{kl})=0, l=1,\cdots,d 
\label{Golden_var_formal}
\eeq
{Define} $\wt{\triangle}:=\sum_{q=1}^{d}L_q^2$. {Then we see that formally $L_j(\wt{\triangle})^{-1}\sum_{k=1}^d L_k  \wt{G}_k^l=L_j(\wt{\triangle})^{-1}\wt{\triangle}\psi^l=L_j\psi^k=\wt{G}^l_j$}. By applying $L_j (-\wt{\triangle}^{-1})$ to \eqref{Golden_var_formal}, followed by adding $\delta_{jl}$ on both sides, the desired expression is achieved
\beq
(\wt{G}_j^l+\delta_{jl})+(1-h)\sum_{k=1}^d L_j (-\wt{\triangle}^{-1})L_k \wt{\chi_2}(\wt{G}_k^l+\delta_{kl})=\delta_{jl},\,j,l=1,\cdots,d.
\label{operator_form}
\eeq

Define a new variable $s:=\frac{1}{1-h}$ and the operator $\wt{B}_{jk}:=L_j(-\wt{\triangle}^{-1})L_k \wt{\chi_2}$ 
It can be shown that $\wt{\B{B}}$ is a self-adjoint bounded linear operator with respect to the inner product
\beqs
<\wt{f},\wt{g}>:=\int_\Omega \wt{\chi_2}(\omega)\wt{f}\cdot \ov{\wt{g}} ,\, \wt{f},\wt{g}\in (L^2(\Omega,\mathcal{F},P))^d
\eeqs
$\|\wt{\B{B}}\|\le 1$. Then the integral equation above becomes 
\beq
\wt{\B{E}}^l(h)=\wt{\B{G}^l}(h)+\B{e}_l=\left(\B{I}+\frac{\wt{\B{B}}}{s}\right)^{-1}\B{e}_l
,\, s=\frac{1}{1-h}{.}
\label{Eh}
\eeq 
Applying the spectral theory of self-adjoint {operators} and taking into account the fact that $s\in[-1,0]$ must be in the resolvent set, the solution is represented in terms of the projection-valued measure $\B{Q}(dz)$ associated with $\wt{\B{B}}$
\beq
\wt{G}_j^l+\delta_{jl}=s\int_0^1 \frac{(\B{Q}(dz) \B{e}_l)_j}{s-z},l,j=1,\dots,d \mbox{ for all } s\in\field{C}\setminus[0,1].
\eeq 
Therefore the effective property in \eqref{eff_perm_h} is given by  the following integral representation formula (IRF)
\beq
\epsilon^*_{kl}(h)=\epsilon_1\left[\delta_{kl}-\int_0^1 \frac{\mu_{kl}(d\xi)}{s-\xi}\right],{\quad \text{ where } h=1-\frac{1}{s}} \label{G_IRF}\\
\mbox{ and } \mu_{kl}(d\xi)=\int_{\Omega} \wt{\chi_2}(\omega)(\B{Q}(d\xi)\B{e}_l)_k P(d\omega)\nonumber
\eeq
In \cite{golden1983bounds-for-effe}, the function $m$ defined in \eqref{m_function} is used in the main theorem, which shows that the diagonal terms in the effective permittivity tensor can be represented in terms of a Stieltjes function with finite positive Borel measures. The main theorem is stated below.
\begin{theorem}[ \cite{golden1983bounds-for-effe}]
Let $s=\frac{1}{1-h}$ and $F_{kl}(s)=\delta_{kl}-m_{kl}(h)$. There exists {(not necessarily positive)} finite Borel measures $\mu_{kl}(d\xi)$ defined on $0\le \xi\le 1$ such that the \textbf{diagonal} elements  $\mu_{kk}(d\xi)$ are positive measures satisfying $F_{kl}(s)=\int_0^1\frac{\mu_{kl}(d\xi)}{s-\xi}$ for all $s\in\field{C}\setminus[0,1]$.
\label{G_thm_1983}
\end{theorem}
This theorem has been generalized for a special case of  polycrystalline materials in \cite{doi:10.1063/1.5127457}. 

Note that $s=\infty$ or equivalently $h=1$ corresponds to the case of $\epsilon_1=\epsilon_2$, i.e. homogeneous media. Once the IRF is obtained, the relation between the moments of the finite measure and the microstructure $\chi_2$ can be established by comparing the coefficients of the Laurent series expansion of the IRF at $s=\infty$ and the Taylor series expansion of $m_{ik}$ at $h=1$, which involves {differentiation} w.r.t. $h$ of the right hand side of  \eqref{m_function}. 
\beqs
\mu_{kl}^{(n-1)}:=\int_0^1 z^{n-1}\mu_{kl}(dz)=\frac{(-1)^{n-1}}{n!} m_{kl}^{(n)}(1), n=1,2,\cdots.  \\
\eeqs

To evaluate $m_{kl}^{(n)}$, the derivatives of $\wt{\B{E}}^l$  are needed. They can be calculated by expanding  \eqref{Eh} near $s=\infty$ ($h=1$) because the functions are analytic there. By comparing the Taylor coefficients at $h=1$ on  the left-hand-side with the Laurent coefficents on the right-hand side, it is clear that
\beq
\frac{1}{n!}\frac{d^n\wt{\B{E}}^l}{dh^n}\Big|_{h=1}=\wt{\B{B}}^n \B{e}_l, n=0,1,2,\dots, l=1,\dots d{.}
\label{BE}
\eeq

This directly implies that the $n$-th moments are related to the $(n+1)$-point correlation functions of the microstructure. Some explicit relations can be derived. For example, differentiating both sides of \eqref{m_function} leads to 
\beqs
\mu_{kl}^{(0)}=m'_{kl}(1)=
\int_\Omega(\wt{\chi_2}(\omega)\wt{E}^l_k(1,\omega)P(d\omega)+
\left[(h-1)\int_\Omega \wt{\chi_2}(\omega)\frac{d\wt{E}^l_k}{dh}P(d\omega)\right]\Big|_{h=1}=p_2 \delta_{kl}
\eeqs
where $p_2$ is the volume fraction of material with permittivity $\epsilon_2$. If the microstructure $\chi_2(\bx,\omega)$ is spatially isotropic, then one can also obtain the exact expression of  $\mu_{kl}^{(1)}$ by differentiating  \eqref{m_function} twice, applying  \eqref{BE} and {using} the kernel function of $(-\triangle)^{-1}$ to obtain 
\beqs
\mu_{kl}^{(1)}=-\frac{1}{2}m''_{kl}(1)=-\int_\Omega \wt{\chi_2}(\omega)\frac{d\wt{E}^l_k}{dh}(1,\omega)P(d\omega)=\frac{p_1 p_2}{d}\delta_{kl}
\eeqs
where $p_1:=1-p_2$.

Instead of in the setting of an unbounded, stationery random media, the Stieltjes IRF for two-phase composites with isotropic constituents can also be derived in a bounded and deterministic setting with various types of boundary conditions. For example, see \cite{zhang2009reconstruction} for the permittivity tensor and \cite{kantor1984improved-rigoro},\cite{Ou2006On-the-integral-elastic} \cite{bruno1993on-the-stiffnes} for elasticity tensors. 

A very nice feature of a Stieltjes IRF like \eqref{G_IRF}  is the separation of influence - the contrast $s$ is in the integrand while all the mircostructural information is encoded in the measure. It has been used to formulate the problem of finding bounds on diagonal terms $\epsilon_{ii}$ as a linear optimization problem over  the set of all measures supported in $[0,1]$ with constraints on the first $n$ moments. Specifically, let $\mc{PM}$ be the set of all positive finite Borel {measures} on $[0,1]$ and consider $m(h):=m_{11}(h)$, its IRF and the set of {measures} with constraints on the first $n$ moments
\beqs
&&1-m(h;\mu)=F(s)=\int_0^1\frac{\mu(d\xi)}{s-\xi}, \, s=\frac{1}{1-h},\,s\in\C\setminus[0,1],\\
&&\mc{M}(a_0,\dots,a_{n-1}):=\left\{\mu\big|  \mu\in\mc{PM}, \mu^{(0)}=a_0,\,\mu^{(1)}=a_1,\,\dots, \mu^{(n-1)}=a_{n-1} \right\}{.}
\eeqs
where $a_j>0$, $j=0,\dots,n-1$ form a positive definite sequence \cite{Akhiezer1965The-classical-m} so they can be the first $n$ moments of a measure. To study the possibles values of the effective properties by mixing two given materials with contrast $h$, consider the following set of possible values
\[
\Lambda(h,a_0,\dots,a_{n-1}):=\left\{1-F(s,\mu)\in \C\big|\,\mu\in\mc{M}(a_0,\dots,a_{n-1}),s\in \field{C}\setminus[0,1] \right\}
\]
Note that not all measures {correspond} to a microstructure {and hence} $\Lambda$ {contains} values that are not achievable by any microstructure. Nevertheless, it contains all possible values of $\ep^*_{11}(h)$. Clearly, with a fixed value of $s\in\field{C}\setminus[0,1]$, $1-F(s,\mu)$ is a bounded linear map on $\mc{M}(a_0,\dots,a_{n-1})$, which is a compact and convex subset of $\mc{M}$ in the topology of weak convergence. Therefore, $\Lambda(h,a_0,\dots,a_{n-1})$ is a compact and convex set in $\field{C}$ and the extreme points of $\mc{M}(a_0,\dots,a_{n-1})$ are weak limits of measures of the form \cite{Karlin1966Tchebycheff-sys} 
\beq
d\sigma(\xi)=\sum_{k=1}^{n} \alpha_k\delta(\xi-\xi_k), \alpha_k\ge 0,\, 1> \xi_1>\xi_2>\cdots>\xi_n\ge 0 \nonumber\\
\mbox{ and } \sum_{k=1}^n \alpha_k \xi_k^j=a_j,\, j=0,1,\dots,n-1{.}
\label{dirac_form}
\eeq
A crucial step in deriving the bounds is to note the structure of \emph{interlacing poles and zeros} of the functions represented by the type of measures in \eqref{dirac_form}. Consider
\beqs 
m(h;d\sigma):= 1-\int_0^1 \sum_{k=1}^{n} \frac{\alpha_k\delta(\xi-\xi_k)}{s-\xi}=1-\sum_{k=1}^n \frac{\alpha_k}{s-\xi_k}{,}
\eeqs
which is a rational function of $s$. Let $s=\rho_k$, $k=1,\dots,n$, $\rho_1\ge \rho_2\ge \cdots \ge \rho_n$ be the zeros of $m(h)$; they must be of real-valued because of the IRF of $m$. Then the following expression is valid
\[
\prod_{k=1}^n \frac{s-\rho_k}{s-\xi_k}=1-\sum_{k=1}^n \frac{\alpha_k}{s-\xi_k} \Rightarrow \alpha_j=-\frac{\prod_{k=1}^n (\xi_j-\rho_k) }{\prod_{k\ne j}{(\xi_j-\xi_k)} } 
\]
The fact that $\alpha_j \ge 0$ for all $j=1,\dots,n$ {and the additional  physical constraint $m(0,d\sigma)>0$} then lead to the interlacing property
\[
0\le \xi_n\le \rho_n\le\cdots \le \xi_1\le \rho_1\le 1
\]
The bounds on $\ep^*_{11}$ can now be derived as follows. Suppose the volume fraction $p_2$ is given. Then the corresponding bounding function $m(h;d\sigma)$ (or extreme points of $\Lambda(h;p_2)$) has the following form because of its zero $s=\xi_1+p_2\le 1$
\beq
m(h;d\sigma)=1-\frac{p_2}{s-\xi_1},\, 0\le \xi_1\le 1-p_2{.}
\label{bounding_mu_1}
\eeq
If a real-valued $h$ is considered and $h\ge 1$ ($s<0$) {then} the bounds of $m(h;\mu)$ are the well-known geometric  bound and the algebraic bound
\[
1-\frac{p_2}{s-(1-p_1)}\le m(h;\mu)\le1-\frac{p_2}{s} \Rightarrow \frac{1}{1-p_2+\frac{p_2}{h}}\le m(h;\mu)\le 1-p_2+hp_2{.}
\] 
If a complex-valued $h$ is considered, then \eqref{bounding_mu_1} provides the bounding curve for the convex hull of  $\Lambda(h;p_2)$ consisting of a cord and an arc on the complex plane.

If the microstructure is assumed to be isotropic, then $\Lambda(h;p_2,\frac{p_1p_2}{d})$ is considered. When $h$ is real and greater than one, the same procedure recovers the well known Hashin-Shtrikman bounds 
\[
1+\frac{p_2}{\frac{1}{h-1}+\frac{p_1}{d}}\le m(h)\le h+\frac{p_1}{\frac{1}{1-h}+\frac{p_2}{dh}}{.}
\]
When $h$ is of complex values, the convex hull of $\lambda(h;p_2,\frac{p_1p_2}{d})$ is bounded by two arcs. More details about the bounding curves can be found in \cite{Cherkaeva1998Inverse-bounds-} and \cite{Milton2002The-Theory-of-C}.

The bounding curves have been utilized for finding bounds on volume fractions of two-component composites from given complex-valued permittivity data \cite{Mcphedran1990Inverse-Transpo}\cite{Cherkaeva1998Inverse-bounds-}. The separation of influence also makes this type of IRF very useful in retrieving microstructural information from given data on the effective parameters \cite{cherkaev2001inverse-homogen}\cite{cherkaev2008dehomogenizatio}\cite{Bonifasi-Lista2009Electrical-impe} \cite{zhang2009reconstruction}  \cite{Orum2012Recovery-of-inc}\cite{golden2011spectral-analys}. The process of reconstructing the measure from data of $\epsilon_{ii}(h)$ is termed \emph{dehomogenization}, whose theoretical foundation is established by E. Cherkaev in \cite{cherkaev2001inverse-homogen}. For applications of IRF in the study of transport in fluid, see \cite{Avellaneda1995Stieltjes-integ,Avellaneda1989Stieltjes-Integ,Avellaneda1991An-integral-rep}\cite{doi:10.1063/1.5127457}.

\subsubsection{IRF for permeability tensors with positive matrix-valued measures}
In Theorem \ref{G_thm_1983}, the Stieltjes function IRF is concluded only for diagonal terms of the effective permittivity tensor $\bep$. A matrix-valued IRF seems to be a more suitable choice for the study of $\bep$. Moreover, the IRF in  Theorem \ref{G_thm_1983} has such a simple form due to the fact that $|sF_{kk}(\sqrt{-1}s)|<M$ for all $s>0$, i.e. $F_{kk}$ decays fast enough along the imaginary axis; this simplifies the IRF for Herglotz-{Nevanlinna} function significantly. In this section, an application that corresponds to a  matrix-valued function that is analytic in $\field{C}^+\cup(0,\infty)$ and does not satisfy the fast decay condition along the imaginary axis is presented. The Herglotz functions that are analytic on one half line  of the real axis have been studied by Kac and Krein in \cite{Kac1974R-functions-ana}, where they are referred to as the Stieltjes function of class $\B{S}$ and $\B{S}^{-1}$. In different references of the literature, the definitions of these two classes sometimes appeared to be interchanged based on whether the singularities are on the left real axis or the right real axis but the representation theorems for each class has also been modified accordingly. Listed below is the definition that best suits the application to be presented in this section {and which is a matrix version of a modification of  Definition {3 in Part I}}. 

\begin{definition}{\cite{dyukarev1986multiplicative,VKStieltjesFunctionsandHurwitz}}
\begin{enumerate}
\item
A matrix-valued function $\B{F}$ {holomorphic} in $\field{C}\setminus (-\infty,0]$ is of class $\B{S}$ if the {following} two criteria are satisfied.

 $\frac{\B{F}(z)-\B{F}^*(z)}{z-\bar{z}}\ge 0$  if  $Im(z)\ne 0$    and   $\B{F}(x)\ge 0$  for $x>0$.

\item
A matrix-valued function $\B{F}$ holomorphic in $\field{C}\setminus (-\infty,0{]}$ is of class $\B{S}^{-1}$ if the {following} two criteria are satisfied.

 $\frac{\B{F}(z)-\B{F}^*(z)}{z-\bar{z}}\le 0$  if  $Im(z)\ne 0$    and   $\B{F}(x)\ge 0$  for $x>0$.
\end{enumerate}
\end{definition}

\begin{theorem}{}
\label{Matrix_Herglotz}
\begin{enumerate}
\item
$\B{F}(z)$ belongs to class $\B{S}$ if and only if there exists a monotonically increasing matrix-valued function $\pmb{\sigma}(t)$ such that the following IRF holds for $z\in \field{C}\setminus(-\infty,0]$
\[
\B{F}(z)=\pmb{A}+\pmb{C}z+\int_{+0}^\infty \frac{z}{z+t} d\pmb{\sigma}(t)
\]
where $\pmb{A}\ge 0$, $\pmb{C}\ge 0$, $\int_{+0}^\infty \frac{1}{1+t} d\pmb{\sigma}(t)<\infty$ and $\pmb{A}+\pmb{C}+\int_{+0}^\infty \frac{1}{1+t} d\pmb{\sigma}(t)>0$. 
\item
$\B{F}(z)$ belongs to class $\B{S}^{-1}$ if and only if there exists a monotonically increasing matrix-valued function $\pmb{\sigma}(t)$ such that the following IRF holds for $z\in \field{C}\setminus(-\infty,0]$
\[
\B{F}(z)=\pmb{A}+\frac{\pmb{C}}{z}+\int_{+0}^\infty \frac{1}{z+t} d\pmb{\sigma}(t)
\]
where $\pmb{A}\ge 0$, $\pmb{C}\ge 0$, $\int_{+0}^\infty \frac{1}{1+t} d\pmb{\sigma}(t)<\infty$ and $\pmb{A}+\pmb{C}+\int_{+0}^\infty \frac{1}{1+t} d\pmb{\sigma}(t)>0$. 
\end{enumerate}
\end{theorem}   
The application considered here is about the transport property of porous materials in the framework of homogenization of periodic media.  The Darcy permeability tensor  $\B{K}^{(D)}$ plays the role of quantifying the transport of fluid in porous media. To study how the microstructure of a porous media influence its Darcy permeability tensor, $\B{K}^{(D)}$ is treated in \cite{Chuan-Bi2021Spectral-Repres} as the limiting case of the two-fluid problem where each isotropic fluid is characterized by its viscosity $\mu_j$, $j=1,2$.The two-fluid problem is originally formulated in \cite{lipton1990darcy} for studying the Stokes equations of flows mixed with tiny stationery bubbles (inclusions). The mixture is assumed to occupy a region $\Omega$ and the tiny inclusions are \emph{periodically} distributed. The tininess of the inclusions leads to the assumption that the side of the periodic cell is $0<\epsilon\ll1$ while the diameter of $\Omega$ is $O(1)$. Let $Q$ denote the unit periodic cell $(0,1)^n, n=2,3$ that contains disjoint parts $Q_1$ and $Q_2$ (inclusion) with interface $\Gamma=\partial Q_1\cap Q_2=\partial Q_2$ such that $Q=Q_1\cup Q_2\cup\Gamma$ and $\Gamma\cap\partial Q=\emptyset$. We assume  inclusions $Q_2$ can distribute in any possible way in a scaled period cell $\epsilon Q$ as long as they do not touch  one another or the periodic cell boundary $\partial(\ep Q)$. For any given $0<\ep\ll 1$, the domain $\Omega$ is covered by a periodic extension of $\ep Q$, denoted by $\wt{\ep Q}$. Note that $\wt{\ep Q}=\wt{\ep Q_1}\cup \wt{\ep Q_2}\cup \wt{\ep \Gamma}$.

For any fixed $\epsilon$, the hosting fluid has constant viscosity $\mu_1$ and occupies region $\Omega_1^\epsilon:=\Omega\cap\wt{\ep Q_1}$ while the inclusions has constant viscosity $\mu_2$ and occupies region $\Omega_2^\epsilon:=\Omega\cap\wt{\ep Q_2}$. The interface between the hosting fluid and the inclusion fluid is denoted by $\Gamma^\epsilon$, i.e.  $\Gamma^\epsilon=\Omega_1^\epsilon\cap\Omega_2^\epsilon$ and $\Omega=\Omega_1^\epsilon\cup\Omega_2^\epsilon\cup \Gamma^\epsilon$. To make this problem amenable to the theory of Herglotz-Nevanllina functions, we assume $\mu_1>0$ and $\mu_2=z\mu_1$ with $z\in\field{C}$. In the tensor notation, it is
\begin{align}
\tilde{\mu}_{ijkl}(\bx;z) &= (\chi_2(\bx) z \mu_1 + \chi_1(\bx) \mu_1)\frac{(\delta_{ik}\delta_{jl}+\delta_{il}\delta_{jk})}{2} \eqdef  \mu(\bx;z) I_{ijkl}, \,z \in \C{.}
\label{def_viscosity}
\end{align}
The two-fluid problem for the unknown fluid velocity $\bu^\epsilon $ and fluid pressure $p^\epsilon$ is formulated as follows
\begin{equation}
\left\{ \begin{split}
\text{div} \left( 2\tilde{\bmu}^{\epsilon}(\bx;z)e(\bu^{\epsilon}) \right) - \nabla p^{\epsilon}  &= -\B{f} \quad \text{ in } \Omega_1^\epsilon\cup\Omega_2^\epsilon \\
\text{div} \bu^{\epsilon} &= 0 \quad \text{ in } \Omega_1^\epsilon\cup\Omega_2^\epsilon\\
\bu^{\epsilon} &= \bo \quad \text{ on } \partial\Omega\\
\llbracket \bu^{\epsilon} \rrbracket=0,\,\bu^{\epsilon} \cdot \bn &= 0 \quad \text{on } {\Gamma}^\epsilon\\
\llbracket \bpi^{\epsilon} \rrbracket\cdot \bn  &= \left( \llbracket \bpi^{\epsilon} \cdot \bn\rrbracket \cdot \bn \right) \bn \quad \text{on } {\Gamma}^\epsilon
\end{split}
\right. 
\label{global_first_form_std_notation}
\end{equation}
where $e(\bu^\ep)=(\nabla \bu^\ep+ \nabla^T \bu^\ep)/2$, $\B{f}$ is a square integrable momentum source, $\llbracket\cdot\rrbracket$ denotes the jump across $\Gamma^\epsilon$, $\bn$ the exterior normal vector of $\Omega_2^\epsilon$, and the stress tensors $\pi(\bu^\epsilon(\bx;z))$ is  defined as (will be denoted $\pi(\bu^\epsilon)$ for brevity)
\begin{equation}
	\pi(\bu^\epsilon)_{ij} = 2\tilde{\mu}^\epsilon_{ijkl} e(\bu^\epsilon (\bx;z))_{kl} - p^\epsilon(\bx;z)\delta_{ij}{.}
	\label{def_total_stress}
\end{equation}
The first jump condition in system \eqref{global_first_form_std_notation} describes the continuity of $\bu^\ep$ across the interface while the second jump condition states that only the normal traction can have a jump across the interface, i.e. the tangential (or shear) traction has to be continuous across the interface.  

Similar to \cite{lipton1990darcy}, it can be shown by using {the}   Lax-Milgram lemma that for every fixed $0<\epsilon\ll1$,\eqref{global_first_form_std_notation}  has a unique solution $\bu^\ep$ and $p^\ep$ is unique up to a constant for all $z\in \C\setminus (-\infty,0]$.  To obtain convergence results, $p^\ep$ is properly normalized to the new pressure $\wt{p^\ep}$ by a procedure described in \cite{lipton1990darcy}. As $\epsilon \to 0$, the solution $\bu^\epsilon$ and  $\wt{p^\ep}$ converge as follows
\begin{equation*}
    \frac{\bu^\epsilon}{\ep^2} \to \bu^0 \quad \text{weakly in } L^2(\Omega)^3,\;\wt{p^{\ep}} \to P \quad \text{strongly in }L^2(\Omega)/\C
\end{equation*}
where $\bu^0$ and $P$ satisfy the homogenized system:
\begin{equation}
    \left\lbrace\begin{split}
        \bu^0 &= -\bK(\nabla P - \B{f})\quad \text{in }\Omega \label{self_K_def}\\
        \text{div } \bu^0 &= 0\quad \text{in }\Omega\\
    \end{split}\right.
\end{equation}
where the {\em self-permeability} $\bK$ is defined as
\begin{equation}
K_{ij}(z) \defeq  \int_{Q} {u^i_j}({\B{x};}z) d\by =  \int_{Q} {\bu^i}({\B{x};}z)\cdot \be_j d\by
\label{permeability_def}
\end{equation}
by the $Q$-periodic, divergence-free solution $\bu^k$ to the following cell problem 
\begin{equation}
\left \{ \begin{split} \text{div}_{\by} \left( 2\tilde{\mu}(\by;z) e( \bu^k) - p^k \bI \right)  + \be_k &= \bo \quad \text{in } Q_1 \cup Q_2\\
\llbracket \bpi \rrbracket\cdot\bn  &= \left( \llbracket \bpi\cdot \bn\rrbracket \cdot \bn \right) \bn \text{ on } \Gamma
\end{split}\right.
\label{cell_prob}
\end{equation}
where $\B{y}$ denotes the coordinates for the unit periodic cell $Q$ and the viscosity tensor in $Q_1\cup Q_2$ is 
\begin{equation}
\tilde{\mu}_{ijkl}(\by;z) = (\chi_2(\by) z \mu_1 + \chi_1(\by) \mu_1)I_{ijkl} = \mu(\by;z)I_{ijkl}{.}
\end{equation}
 
The function space for the cell problem is the Hilbert space 
\begin{align}
H(Q) &:= \left\{ \bv: \bv\in H^1(Q_1\cup Q_2)^3 \biggr\vert \; \text{div}_{\by} \bv = 0,\; \bv \cdot \bn = 0 \text{ in } H^{-\frac{1}{2}}(\Gamma),\right.\nonumber \\
&\qquad \left.{} \llbracket \bv \rrbracket_{\Gamma} = \bo,\; \bv \text{ is } Q \text{- periodic}  \right\} 
\label{space_HQ}
\end{align}
endowed with inner product
\begin{equation}
(\bu,\bv)_{Q} = \int_{Q}  2\mu_1e(\bu) : \overline{e(\bv)} d\by
\label{inner_product_def}
\end{equation}
where the induced norm is denoted by $\norm{\bu}_{Q}^2 :=  (\bu,\bu)_Q$ and the contraction product of two $n\times n$ matrices $\B{A}=\{a_{ij}\}$, $\B{B}=\{b_{ij}\}$ is $\B{A}:\B{B}=\sum_{i,j=1}^n a_{ij}b_{ij}$. Let $\mathcal{R}(Q)$ denote the space of rigid body displacement in $Q$, i.e. $\bu = \bA\by + \bb$ with constant skew-symmetric matrix $\B{A}$ and constant vector $\bb$ in $H(Q)$. Then we have $H(Q)\cap \mathcal{R}(Q)=\{0\}$ because $\bA = 0$ due to the $Q$-periodicity, and $\bu \cdot \bn = 0$ implies $\bb = \bo$. Hence Korn's inequality can be applied to show that the norm $\|\cdot \|_Q$ is equivalent to the $H^1(Q_1\cup Q_2)$-norm. In this setting, it is proved in \cite{Chuan-Bi2021Spectral-Repres} by using {the} Lax-Milgram lemma that the cell problem has a unique solution $\bu(\by,z)\in H(Q)$ and $p(\by, z)\in L^2(Q)/\C$ for all $z\in\field{C}\setminus(-\infty,0]$. Also, $\bu(z)$ is analytic in $\field{C}\setminus (-\infty,0]$ and so is $\bK(z)$. Moreover, $\bK$ in \eqref{permeability_def} can be expressed as the following quadratic form (note the $\overline{z}$ in function $\tilde{\mu}$)
\begin{equation}
    K_{ij}(z)  =  \int_{Q}2  \tilde{\mu}(\by;\overline{z}) \overline{e( \bu^i(z))} : {e(\bu^j(z))}  d\by
 \label{permeability_second_form}  
\end{equation}
and its conjugate transpose $\bK^*:=\overline{\bK^T}$ is
\begin{equation}
    {(K^*)_{ij}}(z) 
     =  \int_{Q}2  \tilde{\mu}(\by;{z}) {e( \bu^j(z))} : \overline{e(\bu^i(z))}  d\by{.}
\end{equation}

We observe the following two properties.
\begin{enumerate}
\item 
Because $K_{ij}(z)-K^*_{ij}(z)=  {2\mu_1}(\overline{z}-z)\int_{Q_2}  \overline{e( \bu^i(z))} : {e(\bu^j(z))}  d\by$, we have
\[
\frac{K_{ij}(z)-K^*_{ij}(z)}{z-\overline{z}}= -{2\mu_1} \int_{Q_2} \overline{e( \bu^i(z))} : {e(\bu^j(z))}  d\by=  -(\bu^j,\bu^i)_{Q_2}=:-A_{ij}
\]
The matrix $\pmb{A}$ is obviously Hermitian.  Furthermore, for any $\pmb{\xi}\in \field{C}^3$, we have $\sum_{i,j=1}^n \overline{\xi_i} A_{ij} {\xi_j}= (\sum_{j=1}^n \xi_j\bu^j,\sum_{i=1}^n \xi_i \bu^i)_{Q_2}\ge 0$. Therefore, 
\[
\frac{\bK(z)-\bK^*(z)}{z-\bar{z}}\le 0 \mbox{ if } Im(z)\ne 0{.}
\]
\item
For $x>0$, 
recall that $K_{ij}(x)= \left(  (\bu^j,\bu^i)_{Q_1}+x  (\bu^j,\bu^i)_{Q_2} \right)$. With a similar argument as before, we have
\[
\bK(x)\ge 0 \mbox{ for }x>0
\]
\end{enumerate}
With these two properties and the fact that $\bK$ is holomorphic in $\field{C}\setminus(-\infty,0]$, we see that $\bK(z)$ is a Stieltjes function of class $\B{S}^{-1}$. Therefore, by Theorem \ref{Matrix_Herglotz} there exists a monotonically increasing matrix-valued function $\pmb{\sigma}(t)$ such that the following integral representation formula holds for $z\in \field{C}\setminus(-\infty,0]$
\[
\bK(z)=\pmb{A}+\frac{\pmb{C}}{z}+\int_{+0}^\infty \frac{1}{z+t} d\pmb{\sigma}(t)
\]
where $\pmb{A}\ge 0$, $\pmb{C}\ge 0$, $\int_{+0}^\infty \frac{1}{1+t} d\pmb{\sigma}(t)<\infty$ and $\pmb{A}+\pmb{C}+\int_{+0}^\infty \frac{1}{1+t} d\pmb{\sigma}(t)>0$. It is proved in \cite{Chuan-Bi2021Spectral-Repres} that there exist two positive numbers $E_1, E_2>1$ such that $\bK(z)$ is analytic in $(-\infty,-2E_1^2)\cup(-\frac{1}{2E_2^2},0)$; $E_1$ and $E_2$ are the extension constant related to $Q_1$ and $Q_2$. It is also shown in \cite{Chuan-Bi2021Spectral-Repres} that  $\bK(\infty)=\bK^{(D)}$, the Darcy permeability of porous media defined in the appendix of \cite{Sanchez-Palencia1980Non-homogeneous} by L. Tartar. Also, $\bK(0)=\bK^{(B)}$, the permeability when the inclusion is bubbles. Therefore, the IRF above can be simplified to 
\[
\bK(z)=\bK^{(D)}+\int_{\frac{1}{2E_2^2}}^{2E_1^2}\frac{1}{z+t} d\pmb{\sigma}(t)
\]
This shows an interesting fact that with $\mu_1\in\R$ fixed, the larger the magnitute of inclusion viscosity $\mu_1 z$ is, the smaller the permeability. To see how the microstructure influences $\bK$, a new variable $s:=\frac{1}{z-1}$ is defined. As a function of $s$, $\bK$ can be shown to be a function of class $\B{S}$ and hence can be expressed with a monotonically increasing matrix-valued function $\pmb{\rho}(t)$ as follows
\beq
\bK(s)=\bK^{(D)}+\int_{\frac{1}{1+2E_1^2}}^{\frac{2E_2^2}{1+2E_2^2}} \frac{s}{s+t}d\pmb{\rho}(t), 
\label{IRF_Ks_prime}
\eeq
which is valid for all $s\in \field{C}\setminus{[-\frac{2E_2^2}{1+2E_2^2},-\frac{1}{1+2E_1^2}]}$. Finally, the link between the moments of measure $\pmb{\rho}$ and the microstructure can be established by expansion at $s=\infty$. See \cite{Chuan-Bi2021Spectral-Repres} for details.

\subsection{Numerical treatment of memory terms in the modeling of materials}\label{subsec:memory}

{In this section, the Stieltjes function structure of the memory kernel is utilized to design an efficient numerical scheme for solving the poroelastic wave equations.}

In the modeling of wave propagation in poroelastic media such as bones and fluid saturated rock or viscoelastic materials such as polymeric fluid, the current state depends on the history of the time evolution of the state from the starting time. As a result, the governing equations contain a time convolution term whose integrand consists of the unknown state function and a pre-described time-dependent kernel function $K$; this convolution integral is referred to as the memory term. For a time domain solver, the presence of the memory terms poses the challenges of a proper time-stepping scheme.  In the literature, it has been handled by storing the history of the solution such as in \cite{masson2010finite} or a proper design of quadrature rules for approximating the memory term in the time domain, e.g. \cite{lu2005wave-field-simu} and the reference therein. For poroelastic wave equations, the memory term appears in the equation of the  generalized Darcy's law, {which relates the pore pressure $p$, the solid velocity $\B{v}$ and $\B{q}$, the fluid velocity relative to the solid, as follows}

\beq\label{Darcy}
{ -\nabla p = \rho_f \frac{\partial \B{v}}{\partial t}+ \left(\frac{\rho_f}{\phi} \right) \check{\pmb{\alpha}} \star \frac{\pr \B{q} }{\pr t}},~\,t>0,
\eeq
where {the matrix} $\check{\pmb{\alpha}}$ is the inverse {Fourier-Laplace} transform of the $\pmb{\alpha}$ {defined in \eqref{tort_perm}}. The physical origin of the memory term is due to the fact that at the micro-scale (scale of the pore size) in the frequency domain, the boundary layer of the viscous pore fluid is frequency dependent, e.g. the viscous skin depth is inversely proportional to the square root of the frequency. In the seminal papers \cite{biot1956theory-high,biot1956theory-of-propa}, M. A. Biot calculated a critical frequency $f_c$, which separates the regime of laminar pore fluid flow from that of turbulent pore fluid flow {and each regime corresponds {to} a different expression of $\pmb{\alpha}(\omega)$.}

However, the discrepancy between the model prediction and experiment observation of wave dissipation has prompted the study of high-frequency corrections that are more general than the one proposed in \cite{biot1956theory-of-propa}. In order to describe these corrections, we need to introduce the physical quantity that encapsulates this complicated viscodynamics, i.e. the dynamic tortuosity {tensor $\pmb{\alpha}(\omega)$} and the dynamic permeability {tensor $\B{K}(\omega)$ with $\omega$ being the frequency}. {For $\omega\ne 0$, $\pmb{\alpha}(\omega)$ and $\B{K}(\omega)$ are related  as follows:}
\beq
\pmb{\alpha}(\omega)=\frac{i \eta \phi}{\omega \rho_f}{ {\B{K}(\omega)}^{-1}}
\label{tort_perm}
\eeq
{with $i:=\sqrt{-1}$.}
 and $\phi$, $\eta$, $\rho_f$ being the volume fraction, the dynamic viscosity and the density of the pore fluid, respectively.

To keep the discussion simple, we consider the isotropic case {$\pmb{\alpha}(\omega)=\alpha(\omega)\B{I}$}.  

One of {the} most widely used correction{s} is derived in \cite{johnson1987theory-of-dynam} by Johnson, Koplik and Dashen (JKD)
\beq
\label{Eq:Tortuosity}
\alpha(\omega)=\alpha_{\infty }+\frac{i\eta\phi}{\omega K_0\rho_f}\left(1-\frac{4i\alpha^2_{\infty}K_0^2\rho_f\omega}{\eta \Lambda^2\phi^2}\right)^{\frac{1}{2}}{=:\alpha^D(\omega)},
\eeq
where $K_0$ is the static permeability, $\alpha_\infty$ the limit of $\alpha$ at infinite frequency and $\Lambda$ a structure parameter related to the surface-to-volume ratio of the pore space; all of these parameters can be measured{. See} \cite{johnson1987theory-of-dynam}. An important ingredient in their derivation is the causality of $K(\omega)$. This is carried out by first considering the gradient force and the fluid velocity field that are time-harmonic with frequency $\omega$, i.e. the one-sided Fourier transform, followed by extending $K(\omega)$ for complex-valued $\omega$.  A function defined on the complex $\omega$-plane is {\bf{causal}} if and only if it is analytic in the upper half plane. Another requirement in the JKD derivation is that a real-valued stimulus $\nabla p e^{-i\omega t}+\nabla \overline{p} e^{i \overline{\omega} t}$ should result {in} a real-valued response. This leads to the symmetry constraint $K(-\overline{\omega})=\overline{K(\omega)}$. According to  \cite{johnson1987theory-of-dynam}, the function in \eqref{Eq:Tortuosity} was chosen because it is the simplest form of functions that are causal and satisfies the aforementioned symmetry constraint. Of course, there is no reason why it has to be in this form. Indeed, in \cite{Charlaix1988Experimental-St}, \cite{Zhou1989First-principle} and \cite{Pride1993Drag-forces-of-}, it is shown that when the cross-section of the pore space varies rapidly enough, the JKD formula in \eqref{Eq:Tortuosity} severely underestimates the imaginary part of the measured dynamic tortuosity for  low frequency $\omega \le \omega_0:=\frac{\eta\phi}{K_0\rho_f\alpha_\infty}$. 

In \cite{avellaneda1991rigorous-link-b}, the {spectrum $\{\epsilon_j\}_{j=1}^\infty$ of the} incompressible Stokes equation {with kinetic viscosity $\nu$, in the pore space
is used to derive the following general  integral representation formula (IRF) for the dynamic permeability} 
\beq
K(\omega)={\frac{\nu}{F} \int_0^{\Theta_1}  \frac{\Theta dG(\Theta)}{1-i\omega \Theta}}, \mbox{ with } G(\Theta)=\frac{\sum_{\Theta_n \le \Theta} b_n^2}{\sum_{n=1}^\infty b_n^2}, F:=\left(\phi\sum_{n=1}^{\infty} b_n^2 \right)^{-1}
\label{perm_IRF_Torq}
\eeq
where {$\Theta_1:=(\nu \epsilon_1)^{-1}<\infty$,} $dG$ is a positive measure with mass 1 and $b_n>0$, $n=1,2,\dots$, ordered in the same order as the non-decreasing eigenvalues, are defined by the orthogonal spectral system of the Stokes equations. This implies the dependence of $K$ on the pore space geometry is encoded in the measure $dG$. {Note that $0<\epsilon_1\le \epsilon_2\le\cdots$,\, $\epsilon_n\rightarrow \infty$ as $n\rightarrow\infty$.}

{The integral representation for $K$ in \eqref{perm_IRF_Torq} shows that $K$ itself is not a Herglotz-Nevanlinna function. However, as } was noted in \cite{ou2014on-reconstructi}, the permeability {in \eqref{perm_IRF_Torq}} can be {related to}  a Stieltjes function with the new variables {$s:=-i\omega$, $\xi:=-\frac{1}{s}$ and }
\beq
R(\xi):=-s \left(\frac{F}{\nu} \right) K(is)= \int_0^{\Theta_1} \frac{\Theta dG(\Theta)}{\xi-\Theta}=:\int_0^{\Theta_1} \frac{d\lambda(\Theta)}{\xi-\Theta}.
\label{R_function}
\eeq
As a result, the tortuosity $\alpha$ can be represented as follows 
\beq
\alpha(\omega)=\frac{\eta\phi}{\rho_f K_0}\left(\frac{i}{\omega}\right) +{\int_0^{\Theta_1}\frac{d\sigma(\Theta)}{1-i\omega \Theta}} \mbox{     for } \omega {\text{ such that}} -\frac{i}{\omega}\in \field{C}\setminus[0,\Theta_1],
\label{T-IRF}
\eeq
where $d\sigma$ is a positive Borel measure that has a Dirac mass at $\Theta=0$ with strength $\alpha_\infty$. It is also shown in \cite{ou2014on-reconstructi} that the JKD tortuosity in \eqref{Eq:Tortuosity} is indeed a special case of \eqref{T-IRF} by finding the corresponding $d\sigma$  for \eqref{Eq:Tortuosity}. In the context of JKD permeability {$K^D(\omega)$}, this IRF result implies that the geometry parameter $\Lambda$ is related to the microstructure as follows 
\beq
\Lambda=\sqrt{\frac{2K_0 \alpha_\infty}{\phi[\frac{\mu_1(d\lambda^D)}{\mu_0^2(d\lambda^D)}-1]}}, 
\eeq
where $\mu_0$ and $\mu_1$ are the zero-th moment and the first moment, respectively, of the {corresponding} measure in \eqref{R_function}
for the JKD permeability. We note that the commonly used formula in the engineering literature is $\Lambda\approx\sqrt{\frac{2\alpha_\infty K_0}{\phi/4}}$. 

According to the theorem proved in \cite{gelfgren1978multipoint}, the multi-point rational approximation $P_{n-1}/Q_n$ of every Stieltjes function $f(z)=\int_a^b\frac{d\lambda(t)}{z-t}$ is itself a Stieltjes function $\int_a^b\frac{d\beta(t)}{z-t}$ with a bounded, non-decreasing $\beta(t)$. Hence the poles of the rational approximation of a Stieltjes function are all simple with positive residue and located in $[a,b]$. Moreover,  It is shown there that  the convergence is  geometrical with order $2n$  in any compact set on the complex plane. We state the theorem that is relevant to the approximation of $\alpha$ here.

{
\begin{theorem}[\cite{gelfgren1978multipoint}]\label{Thm_gelf}
Let $f$ be a  Stieltjes function of the form  $\int_a^b\frac{d\lambda(t)}{z-t}$
and let  $\gamma_k$,  be a set of interpolation points, consisting of $k_1$ real points $x_1,\cdots, x_{k_1}\in \field{R}\setminus [a,b]$, and $k_2$ non-real points $z_1,\cdots, z_{k_2}\in \field{C}\setminus \field{R}$.   Let $P_{n-1}(z)$ and $Q_n(z)$ be polynomials of degree at most $n-1$ and $n$, respectively, with $k_1+k_2+k_3=2n$ such that the following relations are satisfied 
$$
\begin{cases}
f(z) Q_n(z)-P_{n-1}(z)= A(z) \prod_{j=1}^{k_1}(z-x_j) \prod_{j=1}^{k_2}(z-{ z_j})(z- \overline{z}_j)\\
f(z) Q_n(z)-P_{n-1}(z)=B(z) z^{n-k_3-1},
\end{cases}
$$
where $A(z)$, $B(z)$ are analytic in $\field{C}\setminus[a,b]$ and  $B(z)$ bounded at $\infty$. Then for the multi-point rational approximation it holds: 
\begin{enumerate}
    \item 
$[n-1/n]_f(z):=$$\frac{P_{n-1}(z)}{Q_n(z)}=\int_a^b \frac{d\beta(t)}{z-t}$ for some bounded, non-decreasing function $\beta(t)$. 
\item Denote by $\gamma_k$, $k=1,\ldots 2n$ the interpolation points. Fix one interpolation point $v$ and denote
$G_k(z):=\frac{\psi_v(z)-\psi_v(\gamma_k)}{1-\psi_v(z)\overline{\psi_v(\gamma_k)}}$, where $\psi_v(z)=\frac{\sqrt{z-b}-\sqrt{\frac{v-b}{v-a}} \sqrt{z-a}}{\sqrt{z-b}+\sqrt{\frac{v-b}{v-a}} \sqrt{z-a}}$ is a conformal mapping that maps $\field{C}\setminus[a,b]$ onto the interior of the unit circle and  $v$ onto $0$. \\
If $a,b$ are finite numbers then there exists a constant $K_v$, dependent on $f$ and $v$ but not on $n$, such that  for $z\in \field{C}\setminus[a,b]$ it holds
\[ 
\left|f(z)-\frac{P_{n-1}(z)}{Q_n(z)}\right| \le K_v \frac{1}{\max(|z-a|,|z-b|)} \frac{1}{1-|\psi_v(z)| }
\cdot \Pi_{k=1}^{2n} |G_k(z)|.
\]
\end{enumerate}{Moreover}, if for all $n$  the interpolation points $\gamma_k$ are at least at a fixed non-zero distance away from $[a, b]$, then there exists $\triangle_F<1$ such that $|G_k(z)|\le \triangle_F<1$ and hence
\beq
\left|f(z)-\frac{P_{n-1}(z)}{Q_n(z)}  \right| \le K_v \frac{1}{\max(|z-a|,|z-b|)}\frac{1}{1-|\psi_v(z)|}(\triangle_F)^{2n}{.}
\eeq 
Therefore, the approximation converges geometrically in any compact set that does not intersect with $[a, b]$.
\label{N-pade-thm}
\end{theorem}}

Now, we rewrite \eqref{T-IRF} in terms of $\xi$ and rearrange terms to obtain
\beqs
\alpha(\xi)+\frac{\eta \phi}{\rho_f K_0}\xi-\alpha_\infty=\xi {\int_{+0}^{\Theta_1}\frac{d\sigma(\Theta)}{\xi-\Theta}}{.} 
\eeqs
Theorem \ref{Thm_gelf} implies that  
\beqs
\xi {\int_{+0}^{\Theta_1}\frac{d\sigma(\Theta)}{\xi-\Theta}} \approx \xi \sum_{k=1}^n \frac{\rho_k}{\xi-\pi_k}, \mbox{ with } \rho_k>0 \mbox{ and } 0<\pi_k<\Theta{_1}
\eeqs
with error bound 
\beqs
\left|  \xi {\int_{+0}^{\Theta_1}\frac{d\sigma(\Theta)}{\xi-\Theta}} - \xi  \sum_{k=1}^n \frac{\rho_k}{\xi-\pi_k}  \right|\le \frac{|\xi|}{\max(|\xi|, |\xi-\Theta_1|)}\frac{K_v}{|1-\psi_v(\xi)|}(\triangle_F)^{2n}{.}
\eeqs 
Changing the variable back to $s$, it is clear that there exists $r_k>0$ and $p_k>0$ such that 
\[
\alpha(s){\approx}\frac{\eta\phi}{\rho_f K_0}\left(\frac{1}{s}\right) +\alpha_\infty+ \sum_{k=1}^n \frac{r_k}{s-p_k}
\mbox{     for } s \in \field{C}\setminus(-\infty,-\frac{1}{\Theta_1}].
\]
The $r_k$ and $p_k$ can be accurately computed from given nodes $(s_j,\alpha(s_j))_{j=1}^n$ by using two-sided residue approximation with arbitrary precision arithmetics; see \cite{ou2018augmented} \cite{Xie-Ou-Xu-2019} for details. Let $\mathcal{L}$ be the Fourier-Laplace transform {(note that this differs from the Laplace transform in e.g., Section {2.6 in Part I} by a factor $-i$)} 
\beq
\mathcal{L} {f}(\omega):=\int_{\mathbb R^+} f(t) e^{i\omega t} dt =:\hat{f}(\omega).
\label{F-L-Transform}
\eeq
We approximate the transform of the memory term as follows:
\beqs
{\cal{L} }[\check{\alpha }\star \frac{{\pr \B{q}}}{\pr t}](s)&=& \alpha (s)(s {\hat{\B{q}}-\B{q}(0)})\approx \left( \alpha_\infty+\sum_{k=1}^n \frac{r_k}{s-p_k} +\frac{a}{s}  \right) (s{\hat{\B{q}}-\B{q}(0)})\\
&=& \alpha_\infty (s{\hat{\B{q}}-\B{q}(0)})+\left(a+\sum_{k=1}^n  r_k \right) {\hat{\B{q}}} +\left(\sum_{k=1}^n   \frac{r_k p_k }{s-p_k}\right){\hat{\B{q}}}\\
&&-\left(\sum_{k=1}^n  \frac{r_k}{s-p_k} +\frac{a}{s}\right)\,{\B{q}}(0), \mbox{ where } a:=\frac{\eta \phi}{\rho_f K_0}{.}
\label{Lap_trans}
\eeqs
Furthermore, for each of the terms in the sum, since all the singularities $p_k$ are restricted to the left of $s=-\frac{1}{\Theta_1}$, the inverse Laplace transform can be performed exactly by integrating along the imaginary axis (Theorem 9.1.1 in \cite{Dettman1965Applied-Complex})
\[
{\cal{L}}^{-1}\left[\frac{1}{s-p_k}\right](t)=  {  \frac{1}{2\pi i} }\lim_{R\rightarrow\infty} \int_{-i R}^{i R}\frac{1}{\zeta-p_k} e^{\zeta t}d\zeta=  r_k e^{p_k t},\, t>0.
\]
This integral  is calculated by integrating  along $[-Ri, Ri]\cup \{s=Re^{i\theta} |  \pi/2< \theta < 3\pi/2\}$ and applying the residue theorem and letting $R\rightarrow \infty$. As a result, we have for $t>0$
\beqs
\left(\check{\alpha }\star \frac{\pr {\B{q}} }{\pr t}\right) ( \B{x},t)&:=&\int_0^t \check{\alpha }(\tau)\frac{\pr {\B{q}}}{\pr t}( \B{x},t-\tau)d\tau\\
&\approx& \alpha_\infty \frac{\pr {\B{q}}}{\pr t}+ \left(a+ \sum_{k=1}^n  r_k \right) {\B{q}}
-\sum_{k=1}^n  r_k (-p_k) e^{p_k t} \star  {\B{q}} \\
&&-\left( \sum_{k=1}^n r_k e^{p_k t}+aH(t)\right)\,{\B{q}}(0), 
\eeqs
{where $H$ denotes the Heaviside function.
}
Applying a strategy similar to those in the literature \cite{carcione2001wave-fields-in-}, we define the auxiliary variables $\Theta_k$, $k=1,\ldots,n$ such that 

\beq
{\mathcal{\theta}}_k( \B{x},t):= (-p_k) e^{p_k t} \star  {\B{q}}.
\label{theta_def}
\eeq
It can be easily checked that ${\mathcal{\theta}}_k$, $k=1,\ldots,M$, satisfies the following equation:
\beq
\pr_t{\mathcal{\theta}}_k( \B{x},t)=p_k {\mathcal{\theta}}_k( \B{x},t)-p_k {\B{q}}( \B{x},t).
\eeq
Finally, we can approximate the generalized Darcy's law \eqref{Darcy} with the following system that has no explicit memory terms
\begin{empheq}[left=\empheqlbrace]{align}
\pr_t {\mathcal{\theta}}_k (\B{x},t)&=p_k {\mathcal{\theta}}_k (\B{x},t)-p_k {\B{q}}(\B{x},t),\, k=1,\cdots,n\\
-{\nabla p} &= \rho_f \frac{\partial {\B{v}}}{\partial t}+ \left(\frac{\rho_f \alpha_{\infty j}}{\phi} \right) \frac{\pr {\B{q}} }{\pr t}
  + \left( \frac{\eta}{K_{0j}}+\frac{\rho_f}{\phi}\sum_{k=1}^n r_k \right) {\B{q}} \nonumber\\
 &-\left(\frac{\rho_f}{\phi}\right)\sum_{k=1}^n r_k {\mathcal{\theta}}_k -\frac{\rho_f}{\phi}\left( \sum_{k=1}^n r_k e^{p_k t}+a\right)\,{\B{q}}(\B{x},0),\,t>0 
 \label{aug_e}
\end{empheq}

The generalization to anisotropic tortuosity function $\alpha$ is straightforward and has been implemented numerically in  \cite{Xie-Ou-Xu-2019}.
\subsection{Broadband passive quasi-static cloaking}\label{subsec:milton}
{The sum rules for Herglotz-Nevanlinna functions can be applied to explain and quantify the limitations of broadband quasi-static cloaking.}
 In this section, we summarize the results from the paper by Cassier and Milton   \cite{doi:10.1063/1.4989990}. 

{Here the geometry is as follows: $\Omega\subset\field{R}^3$ is an open bounded set, which in this context is thought of  as the whole device. Let then $\mathcal{O}\subset\Omega$ be a bounded simply connected dielectric inclusion with Lipschitz boundary  such that the cloak $\Omega\setminus \overline{\mathcal{O}}$ is open and connected.}

Consider the Maxwell equations for $\B{D}$ (electric induction), $\B{B}$ (magnetic induction), $\B{E}$ (electric field) and $\B{H}$ (magnetic field)
\beq
\partial_t \B{D}-\nabla \times \B{H}=-\B{J},\,\, \partial_t \B{B}+\nabla \times \B{E}=-\B{J}_B,\,\, \nabla\cdot \B{D}=0,\,\, \nabla\cdot \B{B}=0{.} 
\label{Max_eqn}
\eeq
Suppose the  external electric current $\B{J}$ and magnetic current $\B{J}_B$ are absent{;} one has $\B{J}=\B{J}_B=\B{0}$. Let $\epsilon_0$ and $\mu_0$ denote the permittivity constant and the permeability constant of vacuum, respectively. The Maxwell equations are supplemented with the constitutive laws
\beq
\B{D}=\epsilon_0 \B{E}+\B{P}  \mbox{\hspace{0.3in}and\hspace{0.3in}} \B{B}=\mu_0\B{H}+\B{M}
\label{max_const_1}
\eeq
where the electric polarization $\B{P}$ and the magnetic polarization $\B{M}$ are defined by the time convolution with the real-valued electric susceptibility function $\chi_E$ and the magnetic susceptibility function $\chi_M$ as follows
\beq
\B{P}=\epsilon_0\chi_E\star \B{E} \mbox{\hspace{0.3in}and\hspace{0.3in}} \B{M}=\epsilon_0\chi_M\star \B{H}{.}
\label{max_const_2}
\eeq

The functions considered {are} {as follows: for each $\B{x}\in\Omega$ we have ${\chi_E(\B{x},\,\cdot),\, \chi_M(\B{x},\,\cdot\,)}\in L^1(\field{R})$ and
$\B{E},\B{H}\in H^1(\field{R};L^2(\Omega))$.
}

The causality assumption of the material, i.e. $\B{E}(\cdot,t)$ and $\B{H}(\cdot,t)$ cannot influence $\B{D}(\cdot,t')$ and $\B{B}(\cdot, t')$ for $t'<t$, implies {that} $\chi_E(\cdot,t)$ and $\chi_M(\cdot,t)$ are supported in $t\ge 0$. Applying the Laplace-Fourier transform \eqref{F-L-Transform} to \eqref{max_const_1} and \eqref{max_const_2}, the following relations in the frequency domain are obtained{:}
\beq
&&\hat{\B{D}}(\omega)=\epsilon_0(1+\hat{\chi}_E(\omega)) \hat{\B{E}}(\omega)=: \epsilon(\omega)  \hat{\B{E}}(\omega){,}\label{freq_const_1}\\
&&\hat{\B{B}}(\omega)=\mu_0(1+\hat{\chi}_M(\omega)) \hat{\B{E}}(\omega)=: \mu(\omega)  \hat{\B{E}}(\omega){.}
\label{freq_const_2}
\eeq
For real-valued $\omega$, $\epsilon(\omega)$ and $\mu(\omega)$ are the usual dielectric permittivity and the magnetic permeability, respectively. The assumption of $\chi_E(\bx,\cdot),\, \chi_M(\bx,\cdot)\in L^1(\field{R})$ leads to the fact that all the functions involved in \eqref{freq_const_1} and \eqref{freq_const_2} are analytic in $\field{C}^+$ and continuous in the topological closure $cl(\field{C}^+){=\mathbb C^+\cup\mathbb R}$. Moreover, by applying the Riemann-Lebesgue theorem to $\chi_E$ and $\chi_M$, one has $\hat{\chi}_E(\omega)\rightarrow 0$ and $\hat{\chi}_M(\omega)\rightarrow 0$ as $|\omega|\rightarrow \infty$ in $cl(\field{C}^+)$. Therefore, $\epsilon(\omega)\rightarrow \epsilon_0$ and  $\mu(\omega)\rightarrow \mu_0$ as $cl(\field{C}^+)\ni \omega \rightarrow \infty$. The \textbf{passivity} assumption that demands {non-negative } electric/magnetic energy loss is formulated as
\beq
\mathcal{E}_a(t)=\int_{-\infty}^t\int_\Omega \partial_t\B{D}(\B{x},s)\cdot \B{E}(\B{x},s) + \partial_t\B{B}(\B{x},s)\cdot \B{H}(\B{x},s) d\B{x}ds\ge 0, t\in \field{R}{.}
\label{Maxwell_eqs}
\eeq
Then the Plancherel theorem implies that
\beqs
\mathcal{E}_a(\infty)=\frac{1}{2\pi} Re \int_\field{R}\int_\Omega -i\omega \left( \hat{\epsilon}(\B{x}, \omega) |\hat{\B{E}}(\B{x},\omega)|^2 +
\hat{\mu}(\B{x}, \omega) |\hat{\B{H}}(\B{x},\omega)|^2\right) d\B{x}d\omega\ge 0{.}
\eeqs
Since this has to hold for all $\B{E}$ and $\B{H}$,  it must be true that $\omega \im (\epsilon(\omega))\ge 0 $ and  $\omega \im (\mu(\omega))\ge 0 $ for all real-valued $\omega$. These properties of $\hat{\chi}_E$, $\hat{\chi}_M$, $\epsilon(\omega)$ and $\mu(\omega)$ prompt the study of functions $f:cl(\field{C}^+)\rightarrow \field{C}$ that satisfy the following hypotheses
\begin{itemize}
\item
H1: $f$ is analytic in $\field{C}^+$ and continuous in $cl(\field{C}^+)$ (causality).
\item
H2: $f(z)\rightarrow f_\infty>0$ as $|z|\rightarrow\infty$ in $cl(\field{C}^+)$.
\item 
H3: $f(-\overline{z})= \overline{f(z)}, z\in cl(\field{C}^+)$.
\item
H4: $\im f(z)\ge 0$ for all $z\in \field{R}^+$ (passivity){.}
\end{itemize}
{Note that $f$ satisfying hypotheses H1-H4 is not a Herglotz-function, however,  Remark {2 in Part I} implies that the function $v(\omega):=\omega f(\sqrt{\omega})$ is, and this fact will be utilized below}.

The problem of passive, quasi-static cloaking for incident plane waves is formulated in  \cite{doi:10.1063/1.4989990} as follows. Suppose the material in $\mathcal{O}$ has constant permittivity $\epsilon\B{I}$ with $\epsilon>\epsilon_0$ and is non-dispersive (frequency independent) in the frequency range $[\omega_{-},\omega_+]$. The cloak is assumed to have permittivity $\epsilon(\B{x},\omega)$ and {occupies the space $\Omega\setminus \mathcal{O}$} {surrounding} the inclusion $\mathcal{O}$.

It is assumed that the permittivity in $\field{R}^3\setminus \Omega$ is $\epsilon_0 \B{I}$. In the quasi-static case, the time derivatives in \eqref{Max_eqn} are negligible and hence $\B{E}=-\nabla V$ for some scalar potential $V$. Let the incident plane wave be $\B{E}_0$, a uniform field in $\field{R}^3$, which will interact with the device $\Omega$ and the scattered field with potential $V_s$ will be generated. The total potential $V$ is related to the scattered potential $V_s$ by $V(\bx,t)=-\B{E}_0+V_s(\bx,t)$. Then $V_s$ satisfies the equation
\begin{empheq}[left=\empheqlbrace]{align}
\nabla\cdot(\pmb{\epsilon}(\B{x},\omega)\nabla V_s)=\nabla\cdot(\pmb{\epsilon}(\B{x},\omega)-\epsilon_o\B{I})\B{E}_0 \mbox{ in } \field{R}^3,\label{Vs}\\ V_s(\B{x},\omega)={O}(1/|\B{x}|) \mbox{ as } |\B{x}|\rightarrow \infty. \nonumber
\end{empheq}

Because the cloak occupying $\Omega\setminus\mathcal{O}$ is assumed to be passive, the permittivity $\pmb{\epsilon}(\B{x},\omega)$ satisfies the following conditions for almost all $\B{x}\in \Omega\setminus\mathcal{O}$
\begin{itemize}
\item
$\widetilde{\mbox{H}}$1:
$\bep(\B{x},\cdot)$ is analytic on $\field{C}^+$ and continuous on $cl(\field{C}^+)$.
\item
$\widetilde{\mbox{H}}$2:
$\bep(\B{x},\omega)\rightarrow \epsilon_0\B {I}$ as $|\omega|\rightarrow \infty$ in $cl(\field{C}^+)$.  
\item
$\widetilde{\mbox{H}}$3:
$\bep(\B{x},-\overline{\omega})=\overline{\bep(\B{x},{\omega})} \, \forall \omega\in cl(\field{C}^+)$.
\item
$\widetilde{\mbox{H}}$4:
$\im\,{\bep(\B{x},{\omega})}\ge 0 \,\forall \omega\in \field{R}^+$
\item
$\widetilde{\mbox{H}}$5:
${\bep(\B{x},{\omega})}^T={\bep(\B{x},{\omega})},\, \forall \omega\in cl(\field{C}^+)$ (reciprocity principle){.}
\end{itemize}
 
For the well-posedness of \eqref{Vs}, two additional conditions are imposed.
\begin{itemize}
\item
$\widetilde{\mbox{H}}$6:
$\bep(\cdot,{\omega}) \in L^\infty(\Omega\setminus\mathcal{O})\, \forall \omega\in cl(\field{C}^+)$  such that 
$\sup_{\omega\in cl(\field{C}^+)} \|  \bep(\cdot,{\omega})    \|_{L^\infty(\Omega\setminus\mathcal{O})}\le c_1$ with positive constant $c_1$.
\item
$\widetilde{\mbox{H}}$7a:
There exists $c_2(\omega)>0$ and $\gamma(\omega)\in [0,2\pi)$ such that for all $\omega\in \field{C^+}$, one has $| \im (e^{i\gamma(\omega)}  \bep(\B{x},{\omega})\B{E}\cdot\overline{\B{E}} )|\ge c_2(\omega) \|\B{E}\|^2, \,\forall \B{E}\in \field{C}^3$  and {for almost all $\B{x}\in \Omega\setminus\mathcal{O}$.}
\item
$\widetilde{\mbox{H}}$7b:
Let $B(\omega_0,\delta)$ denote the disk centered at $\omega_0$ with radius $\delta$. {For} all $\omega_0\in\field{R}$, there exists $c_3(\omega_0)>0$, $\delta>0$ and $\gamma(\omega_0)\in [0,2\pi)$ such that for all $\omega\in B(\omega_0,\delta)\cap cl(\field{C}^+)$, one has $| \im (e^{i\gamma(\omega_0)}  \bep(\B{x},{\omega})\B{E}\cdot\overline{\B{E}} )|\ge c_3(\omega_0) \|\B{E}\|^2, \,\forall \B{E}\in \field{C}^3$ and {for almost all $\B{x}\in \Omega\setminus\mathcal{O}$.}
\end{itemize}
With these assumptions on $\bep(\B{x},\omega)$,  it is shown in \cite{doi:10.1063/1.4989990} that the potential of the total electric field that satisfies the condition $V(\B{x},\omega)=-\B{E}_0 \cdot \B{x}+{O}(1/|\B{x}|)$ as $|\B{x}|\rightarrow \infty$ is {of the form}
\beq
V(\B{x},\omega)=-\B{E}_0 \cdot \B{x}+\frac{(\pmb{\alpha}(\omega)\B{E}_0) \cdot \B{x}}{4\pi \epsilon_0 |\B{x}|^3}+{O}(1/|\B{x}|^3) 
\eeq 
as $|\B{x}|\rightarrow \infty$   for all $ \omega\in cl(\field{C}^+)\cup\{\infty \}$, where the {complex-valued} $3\times 3$ polarizability tensor $\pmb{\alpha}(\omega)$ is given by
\beq
\pmb{\alpha}(\omega) \B{E}_0=\int_\Omega (\bep(\B{x},\omega) -\epsilon_0\B{I}  )(\B{E}_0-\nabla V_s(\B{x},\omega))\, d\B{x}.
\label{polariz_def}
\eeq
The key point is that $\pmb{\alpha}(\omega)$ describes the leading term of the far-field scattered field generated by the device $\Omega$. Hence the broadband cloaking of the dielectric inclusion $\mathcal{O}$ in the frequency interval $[\omega_-,\omega_+]$ is achieved when $\pmb{\alpha}(\omega)$ vanishes for \emph{all} $\omega\in [\omega_-,\omega_+]$. 
It is proved in \cite{doi:10.1063/1.4989990} that if $\bep(\B{x},\omega)$ satisfies the hypotheses $\rm{\widetilde{H}1}$-$\rm{\widetilde{H}}7b$, then the function $f(\omega):=\pmb{\alpha}(\omega)\B{E}_0\cdot \overline{\B{E}_0}$ satisfies hypotheses H1-H4. Consequently,  
\beq
v(\omega):=\omega f(\sqrt{\omega})=\omega \pmb{\alpha}(\sqrt{\omega})\B{E}_0\cdot \overline{\B{E}_0}
\label{v_def}
\eeq
is a Herglotz-Nevanlinna function {analytic} in $\field{C}\setminus \field{R}_0^+$ and negative in $\field{R}^-$. {Note that $v$ is not a Stieltjes function, since it has {the} "wrong sign" on the negative half line.}   Furthermore, $f(\infty)=\pmb{\alpha(\infty)}\B{E}_0\cdot \overline{\B{E}_0}$ {holds}, which is positive for any non-zero field $\B{E}_0$. This immediately leads to the conclusion that $\pmb{\alpha}(\omega)$ cannot vanish in any interval $[x_-,x_+]$ with $x_-,x_+\in \field{R}^+$ and $x_-\ne x_+$. Because if is does vanish, so does $f$; then the Schwarz reflection principle and the analytic continuation imply $f$ is identically zero in $\field{C}^+$, which contradicts the fact $f(\infty)>0$. Therefore, broadband cloaking is not possible for {a} quasi-static passive cloak. 

We conclude this section by explaining the main ingredients in the derivation of a more refined quantification of the fundamental limits of broadband passive cloaking in quasi-statics presented in \cite{doi:10.1063/1.4989990}. 

Since the polarizability tensor $\pmb{\alpha}(\omega)$ with real-valued $\omega$ is of interest in physics, the Herglotz-Nevanlinna function setting is applied to extract information from the behavior of $\pmb{\alpha}(\omega)$ as a function in $cl (\field{C^+})$ to conclude something useful for its behavior on the positive real line. One important tool for making this connection is {the} sum rule, as stated in Theorem {10 in Part I}{{, which is} applied to the composition with an appropriate window function{. This same technique is } also used in Section {3.1 in Part I}.}

To be able to focus on a finite interval $[-\triangle,\triangle]\subset\field{R}$, $\triangle >0$, the function $h_m$ {is}  defined as
\beq
h_m(z):=\int_{-\triangle}^{\triangle} \frac{dm(\xi)}{\xi-z}
\label{h_function}
\eeq
where $m$ belongs to $\mathcal{M}_\triangle$, the set of finite positive Borel measure supported in $[-\triangle,\triangle]$ such that $m([-\triangle,\triangle])=1$. Obviously, $h_m(z)$ is a Herglotz-Nevanlinna function. By using the theorems in \cite{Bernland_2011} and \cite{doi:10.1063/1.4989990}, the following asymptotic behavior can be concluded 
\[
h_m(z)=-\frac{m(\{0\})}{z}+o\left(\frac{1}{z}\right),\mbox{ as } |z|{\hat\to}0 \mbox{ and } h_m(z)=-\frac{1}{z}+o\left(\frac{1}{z}\right), \mbox{ as }|z|{\hat\to} \infty
\] 
{Since} for any function $f$ that satisfies H1-H4, the corresponding function $v(z):=zf(\sqrt{z})$ is a Herglotz-Nevanlinna function in $\field{C}\setminus \field{R}^+$ and negative in $\field{R}^-$, the composition $v_m(z):=h_m(v(z))$ is again a Herglotz-Nevanlinna function with the following asymptotic {expansion}
\[
v_m(z)=-\frac{m(\{0\})}{f(0)z}+o\left(\frac{1}{z}\right),\mbox{ as } |z|{\hat\to} 0 \mbox{ and } v_m(z)=-\frac{1}{f_\infty z}+o\left(\frac{1}{z}\right), \mbox{ as }|z|{\hat\to} \infty.
\] 
Theorem {10 in Part I} for $n=0$ immediately implies that for any given finite  interval $[x_-,x_+]\subset\mathbb R$ and any $m\in \mathcal{M}_\triangle$, one has 
\beq
\lim_{y\rightarrow 0^+}\frac{1}{\pi} \int_{x_-}^{x_+} \im v_m(x+iy) dx&\le& \lim_{\eta\rightarrow 0^+}\lim_{y\rightarrow 0^+}\frac{1}{\pi} \int_{\eta<|x|<\eta^{-1}} \im v_m(x+iy) dx\nonumber\\
&=&\frac{1}{f_\infty}-\frac{m(\{0\})}{f(0)}\le \frac{1}{f_\infty}{.}
\label{sum_rule_vm}
\eeq

{If the cloak is \emph{lossy}} in the finite band $[\omega_-,\omega_+]$, i.e. $\im{\bep(\B{x},\omega)}$ in \eqref{polariz_def} is not negligible in $[\omega_-,\omega_+]$, {then \eqref{v_def} implies} ${\im} v(x)$ {and hence $\im v_m$} is not negligible for $x\in[\omega_-^2, \omega_+^2]:=[x_-,x_+]$. The choice of  $dm(\xi)=\frac{\pmb{1}_{[-\triangle,\triangle]}(\xi)}{2\triangle}d\xi$  results in $h_m(z)=\frac{1}{2\triangle}\log\frac{z-\triangle}{z+\triangle}$ for all $z\in\field{C}^+$ (branch cut at $\field{R}^+$). Consequently, a lower bound of $\im{h_m(z)}$ can be easily derived to be $\im{h_m(z)}\ge \frac{\pi}{4\triangle}H(\triangle-|z|)$, {where $H$ is the Heaviside function.}

Taking into account the sum rule in \eqref{sum_rule_vm}, one has
\beqs
\lim_{y\rightarrow 0^+}\frac{\pi}{4\triangle} \int_{x_-}^{x_+} H(\triangle-|v(x+iy)|) dx
\le \lim_{y\rightarrow 0^+} \int_{x_-}^{x_+} \im v_m(x+iy) dx \le \frac{\pi}{f_\infty}.
\eeqs
Applying {the} Lebesgue Dominated Convergence theorem to the left side leads to 
$
 \int_{x_-}^{x_+} H(\triangle-|v(x)|) dx \le \frac{4\triangle}{f_\infty}.
$
Finally, letting $\triangle=\max_{x_-\le x\le x_+} |v(x)|$ in the previous inequality leads to the bound 
$
\frac{1}{4}(x_+-x_-)f_\infty\le \max_{x\in[x_-,x_+]} |v(x)|.
$
By identifying $x=\omega^2$, the inequality can be directly translated to the following bound on the polarizability tensor in the frequency band
\[
\frac{1}{4}(\omega_+^2-\omega_-^2) \pmb{\alpha}(\infty)\B{E}_0\cdot\overline{\B{E}_0}\le \max_{\omega\in[\omega_-,\omega_+]} \left|\omega^2  \pmb{\alpha}(\omega)\B{E}_0\cdot\overline{\B{E}_0}\right|{.}
\]
Suppose the cloak has a transparent window in the band $[\omega_-,\omega_+]$, i.e. $\im{\bep(\B{x},\omega)}=0$ for $\omega\in[\omega_-,\omega_+]$ for almost all $\B{x}\in \Omega\setminus\mathcal{O}$.   Then the corresponding $v(z)$ in \eqref{v_def} is real-valued for $z\in[\omega_-^2,\omega_+^2]:=[x_-,x_+]$ because of \eqref{Vs} and \eqref{polariz_def}. In this case, more refined bound{s} can be derived because first of all, $v(z)$ can be extended to be an analytic function in $D:=\field{C}\setminus\{[0,x_-]\cup[x_+,\infty)\}$. By letting the measure used in $h_m$ be a Dirac measure $m=\delta_\zeta$ with $\zeta=v(x_0)$ for some $x_0\in{(}x_-,x_+{)}$
and choosing $\triangle$ so that $-\triangle<\xi<\triangle$, one has $v_{\delta_\zeta}(z)=\frac{1}{v(x_0)-v(z)}$, which is {a} Herglotz-Nevanlinna function with a real pole at $x_0$ and hence must be of multiplicity 1. Therefore, $v'(x_0)\ne 0$. Moreover, this pole must be isolated because $v(x_0)-v(z)$ is analytic in $D$.  Therefore, there must exist a small neighborhood $\mathcal{N}$ around $x_0$ where $v_{\delta_\zeta}$ can be expressed as $v_{\delta_\zeta}(z)=\frac{g(z)}{z-x_0}$ with $g(z)$ analytic in $\mathcal{N}$, $g(x_0)=-\frac{1}{v'(x_0)}$ and real-valued in $\mathcal{N}\cap [x_-,x_+]=:(a,b)$. The sum rule \eqref{sum_rule_vm} implies $\lim_{y\rightarrow 0^+} \int_{a}^{b} \im v_{\delta_\zeta}(x+iy) dx\le \frac{\pi}{f_\infty}$. On the other hand, explicit calculation can be performed using Sokhotski-Plemeli formula to get $\lim_{y\rightarrow 0^+} \int_{a}^{b} \im v_{\delta_\zeta}(x+iy) dx=-\pi g(x_0)=\frac{\pi}{v'(x_0)}$. So one has $0<f_\infty\le v'(x_0)$ for all $x_0\in[x_-,x_+]$. This implies $f_\infty\cdot(x_1-x_2)\le v(x_1)-v(x_2)$ for any $x_1,x_2\in[x_-,x_+]$ such that $x_2<x_1$. Suppose $v(x_2)=0$, then $v(x_1)\ge f_\infty\cdot x_1$ for all $x_1>x_2$. Similarly, if $v(x_1)=0$, then $v(x_2)\le -f_\infty\cdot x_1$ for all $x_2<x_1$ in $[x_-,x_+]$. Therefore, even if $\pmb{\alpha}$ is zero at $\omega_0\in[\omega_-,\omega_+]$, one will have $\pmb{\alpha}(\omega)\le -\pmb{\alpha}(\infty)\frac{\omega_0^2-\omega^2}{\omega^2}$ if $\omega_{-}\le \omega<\omega_0$ and $\pmb{\alpha}(\omega) \ge  \pmb{\alpha}(\infty)\frac{\omega_0^2-\omega^2}{\omega^2}$ if $\omega_0\ < \omega\le\omega_+$. Therefore, one cannot achieve the broadband passive quasi-static cloaking (BPQC) in a transparent window.  

{In conclusion, for the BPQC problem, the Herglotz-Nevanlinna function structure of the function $v(\omega)$ in \eqref{v_def} and the accompanied sum rules not only lead to a proof that BPQC is impossible but also give quantitative limitations of BPQC through providing useful lower bounds.}

\subsection{Hamiltonian structure of Time Dispersive and Dissipative Systems}\label{subsec:Figotin}
Wave dissipation and dispersion appears in many materials. For example, the dynamic tortuosity describes both the dissipation and dispersion mechanism for the poroelastic materials. Also, the dispersive nature of the Maxwell's equations is revealed by the frequency dependent permittivity, permeability and the susceptibility functions in  \eqref{freq_const_1} and \eqref{freq_const_2}. These are examples of linear time dispersive and dissipative (TDD) systems. In a series of work, Figotin and  Schenker \cite{Figotin2005Spectral-Theory}\cite{Figotin2007Hamiltonian-Str}\cite{A.-Figotin2007Hamiltonian-tre} developed a framework for studying the Hamiltonian structure of the linear TDD. Specifically, they consider problems of the following form in a Hilbert space setting
\beq
m\partial_t \B{v}(t)=-i \B{A} \B{v}(t)-\int_0^\infty \B{a}(\tau) \B{v}(t-\tau)d\tau+\B{f}(t)
\label{TDD_sys}
\eeq
where $m>0$ is a positive mass operator
in {a} Hilbert space $H_0$,  $\B{A}$ a self-adjoint operator in $H_0$, $\B{f}(t)\in H_0$ a generalized external force and $\B{a}(t)$ an operator valued retarded friction function {that satisfies $\B{a}(t)=0$ for $t<0$.} The total work done by $f$ is $W=\int_{-\infty}^\infty \re\{(\B{v}(t),\B{f}(t))\} dt$. The term $-i \B{A} \B{v}(t)-\int_0^\infty\B{a}(\tau) \B{v}(t-\tau)d\tau$ is interpreted as the force that $v$ exerts on itself at time $t$ with $-i \B{A} \B{v}(t)$ regarded as the instantaneous term. The time dispersive integral term $\int_0^\infty \B{a}(\tau) \B{v}(t-\tau)d\tau$ is based on two fundamental requirements of \emph{time homogeneity} and \emph{causality}. As a simple example, in \cite{Figotin2005Spectral-Theory}, the authors consider a non-magnetic medium by setting $\B{J}_B=0$, $\B{H}=\B{B}$ (hence $\B{M}=0$), $\nabla\cdot\B{J}=0$, $\nabla\cdot\B{E}=0$  and $\mu_0=\epsilon_0=1$ in \eqref{Max_eqn}-\eqref{max_const_2}. The corresponding TDD for this case is
\beqs
&&\B{v}(\B{x},t)=\begin{pmatrix} \B{E}(\B{x},t) \\ \B{B}(\B{x},t)\end{pmatrix}\in H_0\\
&&H_0:=\{ \B{v}\in L^2(\field{C}^6),\nabla\cdot \B{E}=0=\nabla\cdot \B{B}\}
\\
&&\B{A}=\begin{pmatrix} 0 & i\nabla\times \\ -i \nabla\times &0\end{pmatrix},\, \B{a}(t)=\begin{pmatrix} \partial_t\chi_E\B{I}_{3\times 3}&\B{0}_{3\times 3}\\ 
\B{0}_{3\times 3}&\B{0}_{3\times 3}\end{pmatrix},\,\B{f}(t)=\begin{pmatrix} \B{J} \\ \B{0}\end{pmatrix}
\eeqs
The difficulty of studying the spectral theory a TDD system can be easily seen by considering the time-frequency Fourier transform
$\hat{v}(\omega):=\int_{-\infty}^\infty e^{i\omega t} {v(t)} dt$. {Note that here the Fourier transform is denoted in the same way as the Laplace-Fourier transform in the preceding sections. Then} the TDD system \eqref{TDD_sys} {becomes}
\[
\omega m \hat{\B{v}}(\omega)=(\B{A}- i\hat{\B{a}}(\omega))\hat{\B{v}}(\omega)+{i\hat{\B{f}}(\omega)}=: \hat{\B{A}}(\omega)\hat{\B{v}}(\omega)+i\hat{\B{f}}(\omega){.}
\]
The Kramers-Kronig relations imply that $\hat{\B{A}}$ is non-self-adjoint as long as $\hat{\B{a}}\ne 0$ and hence the eigenvectors of the problem $\omega m \B{e}_\omega=(\B{A}- i\hat{\B{a}}(\omega)) \B{e}_\omega$ are not {necessarily} orthogonal for distinct $\omega$ and may not form a basis for $H_0$. This challenge can be addressed by the method of conservative extension of the TDD system \cite{Figotin2005Spectral-Theory} by first noting that $\hat{\B{A}}$ is not an arbitrary non-selfadjoint operator because the friction operator $\hat{\B{a}}$ has to satisfy certain characteristic properties of physical laws. Also, {as is pointed out in \cite{Figotin2005Spectral-Theory,A.-Figotin2007Hamiltonian-tre,Figotin2007Hamiltonian-Str}, for all DD systems that are \emph{physical}, the frequency dependence of $\hat{\B{a}}$ originates from ignoring its coupling with another system, whose variables are referred to as the {\em hidden degree of freedom} of the TDD system. Hence by finding the coupling system, there will be a Hamiltonian structure of the \emph{extended} system, which consists of  the original TDD system and the coupling system.}

Based on this idea,
{a} coupled system is introduced
\beq
&&m\partial_t \B{v}(t)=-i \B{A} \B{v}(t)-i\pmb{\Gamma} \B{w} (t) +\B{f}(t)\\
&&\partial_t \B{w}(t)=-i\pmb{\Gamma}^{\dagger} \B{v}(t) -i\pmb{\Omega}_1\B{w} (t),\pmb{\Omega}_1 \mbox{ is self-adjoint in } H_1 
\label{conserve_sys}
 \eeq
where $H_1$ {denotes} the Hilbert space of the hidden variables $\B{w}$, $\pmb{\Gamma}: H_1\rightarrow H_0$ the coupling operator between {the hidden variable $\B{w}$ and the observable variable $\B{v}$.}{ Following the notation in the papers reviewed here, $\pmb{\Gamma}^{\dagger}$  denotes the adjoint of $\pmb{\Gamma}$}.   This extended system {should} give the original TDD \eqref{TDD_sys} after eliminating $\B{w}$. Note that the second equation implies $\B{w}=-i\int_0^\infty e^{-i\pmb{\Omega}_1 \tau} \pmb{\Gamma}^\dagger \B{v}(t-\tau)d\tau$. Using this to eliminate $\B{w}$ in the first equation leads to the necessary condition
\beq
\B{a}(t)=\pmb{\Gamma} e^{-i\pmb{\Omega}_1  t } \pmb{\Gamma}^\dagger,\, t>0.
\label{spectral_friction}
\eeq
This spectral representation of the friction function $\B{a}(t)$ {indicates how} the unknowns $(\B{w},\pmb{\Omega}_1, H_1)$ of the desired conservative extension  can be recovered from the given $\B{a}(t)$. Suppose  $\B{a}(t)$ has the general form 
\[
\B{a}(t)=\boldsymbol{\alpha}_\infty\delta(t)+\boldsymbol{\alpha}(t)
\]
{where $\boldsymbol{\alpha}_\infty=\boldsymbol{\alpha}_\infty^\dagger\ge 0$, $\delta(t)$ is the Dirac function and  $\boldsymbol{\alpha}(t)$ is {for every $t\ge0$} a bounded non-negative operator in $H_0$ such that 
}
\beq
\B{0}\le \boldsymbol{\alpha}_\infty\le C \B{I}_{H_0}, C<\infty,\,\text{ and }\, {\sup_{t\ge 0}} \|\boldsymbol{\alpha}(t) \|_{B(H_0)}< \infty.
\label{friction_form}
\eeq
Note that $\boldsymbol{\alpha}_\infty$ 
corresponds to the classic and familiar friction constant. Then \eqref{spectral_friction} {implies} that $\pmb{\Gamma}$ is unbounded if $\boldsymbol{\alpha}_\infty\ne \B{0}$ because $\B{a}(0)=\pmb{\Gamma} \pmb{\Gamma}^\dagger=\boldsymbol{\alpha}_\infty \delta(t)$. Moreover, $\B{a}(t)$ is extended to $t\le 0$ by 
\beq
\B{a}_e(t)=\pmb{\Gamma} e^{-i\pmb{\Omega}_1  t } \pmb{\Gamma}^\dagger,\, -\infty<t<\infty{.}
\label{spectral_friction_ext}
\eeq
Note that $\B{a}_e(-t)=\B{a}_e^\dagger(t)$.  As a result,  the following power dissipation condition must hold 
\begin{eqnarray}
\mathcal{W}_{fr}(\B{v})&:=& -\frac{1}{2}\int_{-\infty}^{\infty}\int_{-\infty}^{\infty} (\B{v}(t), \B{a}_e(t{-\tau}){\B{v}(\tau)})dtd\tau\nonumber\\
&{=}&-\frac{1}{2}\int_{-\infty}^{\infty} \| e^{i\B{\Omega}_1 t} \pmb{\Gamma}^\dagger \B{v}(t) dt\|^2 \le 0 
\label{energy_dissp}
\end{eqnarray}
This power dissipation condition is also a sufficient condition for the existence of a conservative extension for a TDD system \cite{Figotin2005Spectral-Theory}. The construction of the conservative extension involves finding the essentially unique triplet $(H_1,\pmb{\Gamma},\pmb{\Omega}_1)$ from the extended friction operator $\B{a}_e$. Reconstruction in the time domain can be carried out by using  Bochner's theorem. However, due to the unboundedness of {the} operator $\pmb{\Gamma}$ for the general case $\boldsymbol{\alpha}_\infty\ne 0$, the time-domain reconstruction of the triplet involves subtle technicalities for dealing with the unbounded operator{;} see \cite{Figotin2005Spectral-Theory}. On the other hand, as is pointed out also in \cite{Figotin2005Spectral-Theory}, if one formulates the reconstruction problem in the \emph{complex} frequency form, there will be no unbounded operator involved. The intuition is based on the observation that the {Fourier} transform 
of the friction function $\B{a}(t)$ is $\hat{\B{a}}(\zeta)=\boldsymbol{\alpha}_\infty+\hat{\boldsymbol{\alpha}}(\zeta)$, which is an analytic {operator} function. Assume $\B{v}(t)=0$ and $\B{f}(t)=0$ for $t\le 0$. In this setting, the first step is to {Laplace-Fourier} transform the TDD problem \eqref{TDD_sys} to obtain the following linear response equation 
\beq
(\zeta m- \B{A}+i \hat{\B{a}}(\zeta))\hat{\B{v}}(\zeta){=: i {\mf{A}(\zeta)}^{-1}}=i \hat{\B{f}}(\zeta), \, \im{\zeta}>0{.}
\label{admit_def}
\eeq
The power dissipation condition \eqref{energy_dissp} becomes 
\beqs
\re \hat{\B{a}}(\zeta){ :=\frac{\hat{\B{a}}(\zeta)+{\hat{\B{a}}(\zeta)}^\dag}{2}} \ge 0 \mbox{ for } \im \zeta>0
\eeqs
which implies $(\zeta m-[ \B{A}-i \hat{\B{a}}(\zeta)])$ is invertible for $\im \zeta >0$ because $\B{A}$ is self-adjoint. Define the admittance operator  as $\mf{A}(\zeta):=i (\zeta m- \B{A}+i \hat{\B{a}}(\zeta))^{-1}$ for $\im\zeta>0$.  {Note that then both the the operator valued functions $i\hat{\B{a}}(\zeta)$ and $i\mf{A}(\zeta)$ are  Herglotz-Nevanlinna functions. The} equation above can be written as the admittance equation
\[
\hat{\B{v}}(\zeta)=\mf{A}(\zeta) \hat{\B{f}}(\zeta),\,\im{\zeta}>0.
\]

From the definition of the admittance operator $\mf{A}$, it is clear that one can recover $m$, $\B{A}$ and $\hat{\B{a}}$ from $\mf{A}$ as follows:
\[
m^{-1}=-{\lim_{\eta\rightarrow \infty}} \eta {\mf{A}}\mathcal(i\eta),\,\B{A}=-\displaystyle{\lim_{\eta\rightarrow \infty}} {\im}\mf{A}^{-1}(i\eta) ,\, \hat{\B{a}}(\zeta)=i(\zeta m-\B{A})+\mf{A}^{-1}(\zeta)
\]
To identify the spectral decomposition of $\hat{\B{a}}$, one applies the same transform to the conserved system \eqref{conserve_sys} and eliminates $\hat{\B{w}}$ to obtain $\mf{A}(\zeta)=i[\zeta m-\B{A}-\pmb{\Gamma}(\zeta\B{I}_{H_1}-\pmb{\Omega}_1)^{-1}\pmb{\Gamma}^\dagger]^{-1}$. A comparison with  
the admittance operator defined by \eqref{admit_def} reveals that 
\beq
\hat{\B{a}}(\zeta)=i\pmb{\Gamma}(\zeta\B{I}_{H_1}-\pmb{\Omega}_1)^{-1}\pmb{\Gamma}^\dagger{.}
\label{freq_a_spectral_decom}
\eeq
Besides the power dissipation condition $\re \hat{\B{a}}(\zeta) \ge 0$ for $\im \zeta>0$, the condition \eqref{friction_form} implies 
\[
 \hat{\B{a}}(\zeta)=\alpha_\infty+\hat{\alpha}(\zeta),\,\|\hat{\alpha}(\zeta)\|_{B(H_0)} \le \frac{\sup_{t\ge 0}\left\|\alpha(t) \right\|_{B(H_0)}}{ \im \zeta}.
\]
{This implies that Theorem {14 in Part I} can be applied to show the existence of the space of the hidden variables and the operators in the spectral decomposition \eqref{freq_a_spectral_decom}. Below this theorem is formulated as {it is in Theorem 3.13 in \cite{Figotin2005Spectral-Theory}}. Note that $\pmb{\Omega_1}$ and $\pmb{\Gamma}^\dagger$ here correspond to  $A$ and $\Gamma_0$, respectively,  in {Equation (24) of Part I}. }
 
\begin{theorem}
Let  $G(\zeta)$ be a $B(H_0)$-valued analytic function in $\field{C}^+$ with $\im G(\zeta)\ge 0$ for $\zeta\in \field{C}^+$. If $G$ satisfies the growth condition ${\limsup_{\eta\rightarrow +\infty}} \,\eta \|G(i\eta) \|<\infty$, then $G$ has the following representation
\beq
G(\zeta)=\pmb{\Gamma}(\pmb{\Omega}_1-\zeta\B{I}_{H_1})^{-1}\pmb{\Gamma}^\dagger
\eeq
with $\pmb{\Omega}_1$ a self-adjoint operator on a Hilbert space $H_1$ and $\pmb{\Gamma}: H_1\rightarrow H_0$ a bounded map such that 
\[
\pmb{\Gamma} \pmb{\Gamma}^\dagger\B{v}={\lim_{\eta\rightarrow +\infty}}{{-i\eta}} G(i\eta)\B{v} \mbox{ for all } \B{v}\in H_0.
\]
If $H_1$ is minimal in the sense that $\{f(\pmb{\Omega}_1)\pmb{\Gamma}^\dagger\B{v}: f\in C_c(\field{R}),v\in H_0\}$ is dense in $H_1$, then $\{H_1,\pmb{\Omega}_1,\pmb{\Gamma}\}$ is uniquely determined up to an isomorphism.
\end{theorem}
By identifying $G(\zeta)=i\hat{\B{a}}(\zeta)$ in the theorem, we see that \eqref{freq_a_spectral_decom} has a unique solution up to an isomorphism. Therefore, the conservative extension of \eqref{admit_def} exists. In \cite{Figotin2005Spectral-Theory}, the extended system for dielectric Maxwell' equations with a Lorentzian susceptibility function $\chi$ is constructed. The Hamiltonian structure of the TDD can then be studied via the Hamiltonian structure of the conservative extended system.

\section{More general classes of functions}\label{sec:general}
As we have seen in the preceding sections there are a wide range of applications where Herglotz-Nevanlinna functions are a valuable tool. However, there are also many situations where Herglotz-Nevnalinna functions do not suffice. If, for instance, a causal system is not passive then the corresponding analytic function will not have positive imaginary part. Or if a composite material does consist of more than two materials then the corresponding function will depend on more than only one variable. 

On the mathematical side 
the class of Herglotz-Nevanlinna functions has been generalized in several directions. {To give a short overview,} we will concentrate on scalar generalizations only, even if some results do hold for matrix or operator functions as well.
\subsection{Quasi-Herglotz functions}\label{subsec:quasi}

The class of Herglotz-Nevanlinna functions forms a cone (as it is closed under linear combinations with non-negative coefficients) but not a vector space (since multiples with coefficents other then non-negative do not preserve the Herglotz-Nevanlinna property). As also differences of Herglotz-Nevanlinna functions do appear in applications, the class of quasi-Herglotz functions has been introduced, see \cite{IVANENKO2020Quasi-Herglotz-}. For more details concerning this section see \cite{LugerNedic-quasi}.  

\begin{definition}
A function $q:\mathbb C\setminus\mathbb R\to\mathbb C$ is called a quasi-Herglotz function if it can be written in the form $q=h_1-h_2+i(h_3-h_4)$, where $h_i$ for $1=1,2,3,4$ are Herglotz-Nevanlinna functions (symmetrically extended to the lower halfplane). 
\end{definition}

\begin{example}
Every analytic function $q:\mathbb C^+\to \mathbb C$  with $\im q(z)\geq -c$ for some $c>0$ is a quasi-Herglotz function, since it can be written in the form $q(z)= (q(z)+ic)-ic $, with both $q+ic$ and $ic$  Herglotz-Nevanlinna functions. 
\end{example}

It is  obvious from the definition that this class coincides with all linear combinations of Herglotz-Nevanlinna functions. Hence these functions also can be characterized in terms of an integral representation, however, with complex measures. Recall that complex measures by definition are finite, {see e.g., \cite[Chapter 6]{Rudin},} and hence the representation of the form {of Equation (2) in Part I}{ is used}.
\begin{proposition}\label{pr:quasi}
A function $q$ is a quasi-Herglotz function if and only if there exist real numbers $a$ and $b$ and a complex measure $\sigma$ such that
\begin{equation}\label{eq:intrepquasi}
f(z)=a+bz+\int_{\mathbb R}\frac{1+\xi z}{\xi-z}d\sigma(\xi). 
\end{equation}
Moreover, $a,b$, and $\sigma$ are unique with this property. 
\end{proposition}
Note that quasi-Herglotz functions by definition are defined both in the upper and the lower- halfplane. In contrast to Herglotz-Nevanlinna functions the values in one halfplane do not determine the values in the other. 
\begin{example}
The functions 
$$
q_1(z)=\left\{\begin{array}{rc}
  i   &  \im z>0\\
  -i  &  \im z<0
\end{array}\right.
\quad\text{ and }\quad 
q_2(z)=\left\{\begin{array}{cc}
  i   &  \im z>0\\
  i  &  \im z<0
\end{array}\right.
$$
do coincide in the upper halfplane, but not in the lower. Both are quasi-Herglotz functions as they can be written in the form \eqref{eq:intrepquasi} with $(a_1,b_1,\mu_1)=(0,0,\frac1\pi d\lambda_{\mathbb R})$ for $q_1$  (as in Example {2 of Part I}) and $(a_2,b_2,\mu_2)=(i,0,0)$ for $q_2$.
\end{example}
Considering the difference of the two functions in the example above shows that there are non-trivial quasi-Herglotz functions vanishing identically in one half-plane. All these have been characterized in \cite{LugerNedic-quasi}. 

Given a function, neither the definition nor the characterization in Proposition \ref{pr:quasi} are practical to check whether it is a quasi-Herglotz function or not. But these functions can also be characterized by their analytic properties.

\begin{theorem}\label{thm:quasi_Tumarkin_Vladimirov}
Let $q \colon \mathbb C\setminus\mathbb R \to \mathbb C$ be a holomorphic function. Then $q$ is a quasi-Herglotz function if and only if the function $q$ satisfies, first, a growth condition, namely, that there exists a number $M \geq 0$ such that for all $z \in \mathbb C\setminus\mathbb R$
\begin{equation}
    \label{eq:Vladimirov_growth}
    |q(z)| \leq M \frac{1+|z|^2}{|\im z|},
\end{equation}
 and, second, the regularity condition
\begin{equation}\label{eq:Tumarkin_condition}
\sup_{y \in (0,1)}\int_{\mathbb R}\big|q(x+i y)-q(x-i y)\big|\frac{d x}{1+x^2} < \infty.
\end{equation}
\end{theorem}
An important subclass are real quasi-Herglotz functions{;}  these are real linear combinations of Herglotz-Nevanlinna functions or, equivalently, functions that admit an integral representation \eqref{eq:intrepquasi} with a signed (i.e., real) measure $\sigma$. It can be shown that these functions are exactly those, which are symmetric with respect to the real line, i.e., $q(\overline z)=\overline{q(z)}$.

It can be noted that quasi-Herglotz functions also appear naturally when dealing with Herglotz-Nevanlinna functions only, namely as the off-diagonal elements of matrix-valued Herglotz-Nevanlinna functions. 
\subsection{Generalized Nevanlinna functions}
In the definition of Herglotz-Nevanlinna functions the sign of the imaginary part is required to be positive. However, using the integral representation, it can be shown that this is equivalent to the requirement that the so-called Nevanlinna kernel
\begin{equation}\label{eq:kernel}
 N_f(z,w):=\frac{f(z)-\overline{f(w)}}{z-\overline w} 
\end{equation}
is {positive.} 
{{Recall, that a kernel $N_f(z,w)$ is said to be {positive definite} if for any choice of $N\in\mathbb N$ and $z_1,\ldots,z_N\in\mathcal D$ the matrix
$$
\left(N_f(z_i,z_j)\right)_{i,j=1,\ldots N}
$$
is positive semidefinite.}

{This view leads to the {following} generalization by considering kernels with finitely many negative squares, \cite{KreinLanger1977}.
A   kernel is said to  have $\kappa$ negative squares  if every such matrix above
 has at most $\kappa$ negative eigenvalues and $\kappa$ is minimal with this property.}
\begin{definition}
A function $q:\mathcal D\subset\mathbb C^+\to\mathbb C$ is called a generalized Nevanlinna function if it is meromorphic in $\mathbb C^+$ and the Nevanlinna kernel $N_q$ has finitely many negative squares. If this number is $\kappa$ then  $q\in\mathcal{N}_\kappa$. 
\end{definition}

 Generalized Nevanlinna functions do also admit an integral representation, but it is much more involved than {Equation (1) in Part I}{;} see \cite[Satz 3.1.]{KreinLanger1977}.

The operator representation, however, carries over quite naturally. The only difference {compared} to {Equation (23) in Part I} is that in this case the space is not a Hilbert space, but a Pontryagin space, that is a vector space equipped with an indefinite inner product, such that any non-positive subspace is finite dimensional.

\begin{theorem}\label{thm:oprepgen}
A function $q$ is a generalized Nevanlinna function if and only if there exist a Pontryagin space $\mathcal K$, a self-adjoint linear relation $A$, a point $z_0\in\mathbb C^+$ and an element $v\in\mathcal K$ such that 
\begin{equation}
q(z)=\overline{q(z_0)} +(z-\overline{z_0})\left[(I+(z-z_0)(A-z)^{-1})v,v\right]_{\mathcal K}.
\end{equation}
Moreover, if $\mathcal K=\overline{span}\{(I+(z-z_0)(A-z)^{-1})v: z\in\varrho(A)\}$, then the representation is called minimal. In this case $\mathcal K$ has $\kappa$ negative squares if and only if $q\in\mathcal{N}_\kappa$  and the representation is unique up to unitary equivalence.
\end{theorem}
The conditions on the function $q$ for simplified representations are literally  the same as before and Theorem {7 in Part I} holds for generalized Nevanlinna functions as well. 

From Theorem \ref{thm:oprepgen} and the spectral properties of self-adjoint relations in Pontryagin spaces it follows directly that a generalized Nevanlinna function $q\in\mathcal N_\kappa$ has at most $\kappa$ poles in the upper half plane $\mathbb C^+$, and there are at most $\kappa$ real points $\alpha\in\mathbb R$ (including $\infty$) where it does not hold that  $\lim_{z\hat\to\alpha}(\alpha-z)q(z)$ exists as a non-negative number. These exceptional points (non-real and real) are exactly those eigenvalues, for which the corresponding eigenspace is not a positive subspace. These points are called generalized poles not of positive type. Generalized zeros   not of positive type of $q$ are by definition the generalized poles not of positive type of the   inverse function $\hat q(z):=-\frac1{q(z)}$ (which   belongs to the same class $\mathcal {N}_\kappa$ as $q$).  The importance of these points becomes visible in the following characterization{;} see \cite{DijksmaLangerLugerShondin2000} and also \cite{DerkachHAssideSNoo1999}.

\begin{theorem}\label{thm:fac}
A function $q$ is a generalized Nevanlinna function if and only if there is a rational function $r$ and a Herglotz-Nevanlinna function $f$ such that 
\begin{equation}
q(z)=\overline{r(\overline z)}f(z) r(z).     
\end{equation}
In this case $q\in\mathcal{N}_\kappa$ if and only if $\deg r=\kappa$. Moreover, $r$ is of the form $r(z)=\dfrac{\prod_{i=1}^\ell(z-\alpha_i)}{\prod_{j=1}^m(z-\beta_j)}$, where  $\alpha_i$ are the generalized zeros not of positive type and $\beta_j$ are the generalized poles not of positive type of $ q$ and $\kappa=\max\{\ell,m\}$.
\end{theorem}
Generalized Nevanlinna functions with polynomial $r$ appear for instance in connection with Sturm-Liouville operators with strongly singular potentials or, more abstractly, with strongly singular perturbations of self-adjoint operators in Hilbert spaces, see e.g., \cite{DijksmaLangerShondinZeinstra2000, DijksmaKurasovShondin2005, KurasovLuger2011}.

\begin{remark}\label{rem_randgen}
 A generalized Nevanlinna function $q$ does satisfy  $\lim\limits_{z\hat\to x_0} \im q(z)\geq0$ (as a finite number or $+\infty$) for all but finitely many $x_0\in\mathbb R\cup\{\infty\}$, cf.,  Remark {2 in Part I}.
\end{remark}

\begin{remark}
Also matrix- and operator valued generalized Nevanlinna functions can be defined via a corresponding kernel condition. The operator representation in Theorem {13 of Part I} carries over with the same changes as for scalar functions. The factorization, however, becomes a lot more delicate. The first part holds with rational factors $R(z)$ and $R(\overline z)^*$, but these are not of a comparably simple form, in particular, since generalized poles and zeros can be at the same points. For details see \cite{Luger2002,Luger2003}.
\end{remark}
\subsection{Pseudo-Nevanlinna functions}
\begin{definition}
A function $g$ is called Pseudo-Nevanlinna if it can be written as the quotient of two bounded analytic functions (defined in $\mathbb C^+$) and satisfies $\lim\limits_{z\hat\to x_0} \im g(z)\geq0$
for almost all  $x_0\in\mathbb R$.
\end{definition}
Note that every Herglotz-Nevanlinna function belongs to this class since it can be written as a fractional linear transformation of a function mapping $\mathbb C^+$ into the closed unit disc $\overline{\mathbb D}$. Moreover, by Remark \ref{rem_randgen}  generalized Nevanlinna functions are also pseudo-Nevanlinna functions.  

It has been shown in \cite{DelsarteGeninKamp1986a,DelsarteGeninKamp1986b}  that pseudo-Nevanlinna functions can also be characterized via a factorization, extending Theorem \ref{thm:fac}. To this end one needs to introduce the so-called density functions{;} these are particular pseudo-Nevanlinna functions, which are non-negative (or $\infty$) on the real line.

\begin{theorem}
A function $g$ is a pseudo-Nevanlinna function if and only if there exists a density function $I$ and a Herglotz-Nevanlinna function $g_0$ such that $g(z)=I(z)g_0(z)$.
\end{theorem}
To be precise, in \cite{DelsarteGeninKamp1986a} Pseudo-Caratheodory functions are studied{;} these are corresponding generalizations of Caratheodory functions, i.e., holomorphic functions mapping the open unit disk $\mathbb D$ to the closed right halfplane $\mathbb C_+\cup i\mathbb R$. However, due to the topic of this text here we consider  the corresponding version for the upper halfplane. 

The introduction of Pseudo-Caratheodory functions was motivated by  problems arising in digital signal processing and in the theory of circuits and systems.

\subsection{Functions in several variables}
For analytic functions in one variable in the upper halfplane $\mathbb C^+$ there are two equivalent ways of defining Herglotz-Nevanlinna functions, either by the requirement that $\im  f(z) $ has to be non-negative or that the Nevanlinna kernel $N_f(z,w)$ has to be positive semidefinite. When considering functions in several variables, however, the generalizations along these two ways lead  into different directions, in one case the functions are represented by some kind of resolvents, in the other case by integrals.

In the following we use the notation $\vec z=(z_1,z_2,\ldots, z_n)$ and consider analytic functions $H:(\mathbb C^+)^n\to\mathbb C$, that is $H$ is analytic in each variable $z_j$ for $j=1,\ldots,n$. 

\subsubsection{Loewner functions}

\begin{definition}
A function $H:(\mathbb C^+)^n\to\mathbb C$ is called {a} Loewner function if it is holomorphic and there exist positive semidefinite kernels $A_1,\ldots,A_n$  on $(\mathbb C^+)^n$ such that
\begin{equation}
H(\vec z)-\overline{H(\vec w)}= \sum_{j=1}^n(z_j-\overline{w_j})A_j(\vec z,\vec w)  
\end{equation}
for all $\vec z,\vec w\in(\mathbb C^+)^n$. 
\end{definition}

Loewner functions with $n=1$ are exactly Herglotz-Nevanlinna functions in one variable. 

For $n>1$ these functions have been characterized  in different ways, in particular, as operator monotone functions{;} see \cite{AglerMcCarthyYoung2012}. It has also been shown that functions in this class admit an operator representation, \cite{AglerTully-DoyleYoung2016}. As an example we give one result, corresponding to Theorem {7 in Part I} with $s=0$, in order to show the {flavor} of such representations. 

\begin{theorem}
A function $H:(\mathbb C^+)^n\to\mathbb C$ is a Loewner function satisfying 
$$
\liminf_{y\to\infty}y|\im H(iy, \ldots,iy)|<\infty
$$
if and only if there exists a Hilbert space $\mathcal H$, a self-adjoint operator $A$ in $\mathcal H$, positive  contractions $Y_1,\ldots,Y_n$ with $Y_1+\ldots+Y_n=I_{\mathcal H}$, and an element $v\in\mathcal H$ such that
$$
H(\vec z)=\left((A-z_1Y_1-\ldots-z_nY_n)^{-1}v,v\right)_{\mathcal H}.
$$
\end{theorem}

For Loewner functions, transfer function realizations have {also }been established{;} see \cite{BallKaliuzhnyi-Verbovetskyi2015}.

\subsubsection{Herglotz-Nevanlinna functions.}
The other way of considering several variables leads to the following, more general,  definition. 
\begin{definition}
A function $F:(\mathbb C^+)^n\to\mathbb C$ is called Herglotz-Nevanlinna  function if it is holomorphic and $\im F(\vec z)\geq0$
for all $\vec z\in(\mathbb C^+)^n$.
\end{definition}
It can be shown that not only for $n=1$ but also for $n=2$ the class of Herglotz-Nevanlinna functions do coincide with the class of Loewner functions. However, it is known that this is not true for $n>2${. If $n>2$,} then every Loewner function is a Herglotz-Nevanlinna function, but not conversely. 

For the larger class of Herglotz-Nevanlinna functions in several variables a characterization via an  integral representation has been shown. In order to formulate this result we introduce the following notation. For   $\vec{z} \in (\mathbb C^+)^{n}$ and $\vec{t} \in \mathbb R^n$ define
\begin{equation}\label{eq:kernel_Nvar}
K_n(\vec{z},\vec{t}) := i\left(\frac{2}{(2i)^n}\prod_{\ell=1}^n\left(\frac{1}{t_\ell-z_\ell}-\frac{1}{t_\ell+i}\right)-\frac{1}{(2i)^n}\prod_{\ell=1}^n\left(\frac{1}{t_\ell-i}-\frac{1}{t_\ell+i}\right)\right), 
\end{equation}
which for $n=1$ coincides with the integrand in {Equation (1) of part I}.

Moreover, we say that a Borel measure $\mu$ on $\mathbb R^n$ satisfies the Nevanlinna condition if for all $\vec{z} \in (\mathbb{C}^{+})^n$ and all indices $\ell_1,\ell_2 \in \{1,2,\ldots,n\}$ with $\ell_1 < \ell_2$ it holds 
\begin{equation}
\label{eq:measure_Nevan}
\int_{\mathbb R^n}\frac{1}{(t_{\ell_1} - z_{\ell_1})^2(t_{\ell_2} - \bar{z_{\ell_2}})^2}\prod_{\substack{j=1 \\ j \neq \ell_1,\ell_2}}^n\left(\frac{1}{t_j - z_j} - \frac{1}{t_j - \bar{z}_j}\right) d\mu(\vec{t}) = 0.
\end{equation}

Then the following theorem holds{; see} \cite[Theorem 4.1]{LugerNedic2019}{.}

\begin{theorem}\label{thm:intRep_Nvar}
A function $F:(\mathbb C^+)^n\to\mathbb C$ is a Herglotz-Nevanlinna function if and only if there exist a real number $a \in \mathbb R$,  a vector $\vec{b} \in [0,\infty)^n$
and a positive Borel measure $\mu$ on $\mathbb R^n$ satisfying the Nevanlinna condition and with  
$\int_{\mathbb R^n}\prod_{\ell=1}^n\frac{1}{1+t_\ell^2}d\mu(\vec{t}) < \infty
$
such that
\begin{equation}\label{eq:intRep_Nvar}
F(\vec{z}) = a + \sum_{\ell=1}^nb_\ell z_\ell + \frac{1}{\pi^n}\int_{\mathbb R^n}K_n(\vec{z},\vec{t})d\mu(\vec{t}){.}
\end{equation}
 Furthermore, for a given function $F$, the triple of representing parameters $(a,\vec{b},\mu)$ is unique.
\end{theorem}

Note that for $n=1$ the Nevanlinna condition is satisfied for every measure (which satisfies the necessary growth condition) and hence this theorem becomes Theorem {1 in Part I}. However, for $n>1$ this condition is rather restrictive and measures satisfying it are rather particular. {For example such a measure cannot have finite total mass and hence, in particular, not compact support. There are also other geometric restrictions on the support{;}} see \cite{LugerNedic2021}. 

\section{Summary}
{In this {two-part} survey paper, we start with introducing the various forms of Herglotz-Nevanlinna functions. These definitions are very simple to describe but imply many properties that are physically relevant.  As can be seen from the diverse set of applications presented here, the Herglotz-Nevanlinna functions indeed provide a clear { mathematical} language for {describing} important physical properties such as passivity and causality.} From there, a rigorous analysis can be applied to derive useful properties of these physical system such as the bounds of effective properties of materials or to suggest a way to fabricate materials of desired properties through exploiting the links between some simple forms of Herglotz-Nevanlinna functions and laminated microstructure structure or  their links with some simple circuits. Numerically, the Herglotz-Nevanlinna function theory points a way for approximating memory terms that appear very often in a dispersive system but whose description are given only in the frequency domain. Also, it can provide a framework for {studying} the spectral theory of a TDD system. 

{ {Some very interesting results which involve yet another variation of Herglotz function can be found in the paper by Cassier, Welters and Milton  \cite{cassier2016analyticity}}, where the Dirichlet-to-Neumann(DtN) map for the time-harmonic Maxwell's equations of a two-component {composite} is proved to be a Herglotz-Nevanlinna function of the variable $(\omega\mu_1, \omega\mu_2,\omega\epsilon_1,\omega\epsilon_2)$ in $(\C^+)^4$, which represents the electromagnetic properties of the isotropic constituent materials. To extend the result to the general case of anisotropic constituents, which can be spatially piecewise-constant or continuous, the authors define the class of  Herglotz-Nevanlinna functions on an open, connected and convex set of \emph{matrices with positive definite imaginary parts}. To preserve the Herglotz function structure, they use the \emph{trajectory method} \cite{bergman1993hierarchies-of-} \cite[Section 18.6]{Milton2002The-Theory-of-C} to define a trajectory $s(\omega)$ that maps $\omega$ to the matrix-valued $(\omega\pmb{\mu}_1, \omega\pmb{\mu}_2,\omega\pmb{\epsilon}_1,\omega\pmb{\epsilon}_2)$ and show that the DtN map is a Herglotz-Nevanlinna function along each trajectory. The  implication of this result in electric-impedance-tomography is yet to be discovered.}

With all the applications where Herglotz-Nevanlinna functions have been successfully applied, there are still many open problems that demand further investigations. For example, the IRF for a three-phase dielectric {composite} has been derived in \cite{GOLDEN1986333} by using the theory of Herglotz-Nevanlinna functions  of {two complex variables} \cite{koranyi1963holomorphic-fun}. Also, for the purpose of separating the influence of contrasts and microstructure, a two-parameter IRF has been derived for composites of isotropic elastic materials in \cite{ou2012two-parameter-i} using the results in \cite{koranyi1963holomorphic-fun}. However, in these applications, the relations between the moments and the microstructure become much more complicated. Besides, the characterization of extreme sets of measures of two variables are no longer just {weak} limits of sum of Dirac measures. Also, suppose a set of measurements from a causal and passive system is polluted by noise{;} how can one design a filter to recover the 'nearest' Herglotz-Nevanlinna function that best {represents} the measured data? With the advance of material sciences, there are materials with negative indices and systems that emit energy{;} how should the Herglotz-Nevanlinna function be generalized accordingly? As is described {in Section \ref{sec:general}}, there have been some generalization on the pure mathematics side. We believe {that progress on generalizations} can be sped up by collaboration and communication between mathematicians and {researchers} in various fields of materials sciences through the availability of a set of common mathematical languages and {notation}. 

\bibliographystyle{plain}
\bibliography{library.bib}
\end{document}